\documentclass{jfm}
\usepackage{graphicx}
\usepackage{epstopdf, epsfig}
\usepackage{natbib}
\usepackage{lineno}
\usepackage{cleveref}
\usepackage{subfigure}
\usepackage{textcomp}
\usepackage{setspace}
\usepackage{float}

\shorttitle{Turbulent heat exchange between water and ice}
\shortauthor{E. Ramudu, B. H. Hirsh, P. Olson, and A. Gnanadesikan}

\title{Turbulent heat exchange between water and ice at an evolving ice--water interface}

\author{Eshwan~Ramudu,
  \corresp{\email{eramudu@jhu.edu}}
  Benjamin~Henry~Hirsh,
  Peter~Olson,
 \and Anand~Gnanadesikan}

\affiliation{Department of Earth and Planetary Sciences, Johns Hopkins University, Baltimore, MD~21218, USA}

\begin{document}

\maketitle 

\begin{abstract}
 We conduct laboratory experiments on the time evolution of an ice layer cooled from below and subjected to a turbulent shear flow of warm water from above. Our study is motivated by observations of warm water intrusion into the ocean cavity under Antarctic ice shelves, accelerating the melting of their basal surfaces. The strength of the applied turbulent shear flow in our experiments is represented in terms of its Reynolds number $\Rey$, which is varied over the range $2.0\times10^3 \le \Rey \le 1.0\times10^4$. Depending on the water temperature, partial transient melting of the ice occurs at the lower end of this range of $\Rey$ and complete transient melting of the ice occurs at the higher end. Following these episodes of transient melting, the ice reforms at a rate that is independent of $\Rey$. We fit our experimental measurements of ice thickness and temperature to a one-dimensional model for the evolution of the ice thickness in which the turbulent heat transfer is parameterized in terms of the friction velocity of the shear flow. The melting mechanism we investigate in our experiments can easily account for the basal melting rate of Pine Island Glacier ice shelf inferred from observations. 
\end{abstract}


\newcommand{\D}{\mathrm{d}}
\newcommand\solidrule[1][0.6cm]{\rule[0.5ex]{#1}{.4pt}}
\newcommand\solidrulehalf[1][0.3cm]{\rule[0.75ex]{#1}{.4pt}}
\newcommand\dashedrule{\mbox{%
		\solidrule[1.5mm]\hspace{1.5mm}\solidrule[1.5mm]\hspace{1.5mm}\solidrule[1.5mm]}}

\section{Introduction}\label{sec:intro}
The exchange of heat across the turbulent boundary layer at the ice-ocean interface governs the rate at which sea ice and ice shelves melt or grow in response to changes in ocean properties. The estimation of this heat exchange varies across observational and modelling studies. In order to understand and explain the evolution of sea ice and ice shelves more accurately, it is important to constrain this process. 

Antarctica is surrounded by ice shelves, thick floating sheets of ice that extend from the coastline onto the ocean surface. They play a critical role in the mass balance and dynamics of Antarctica's terrestrial ice by serving as a buttress at the coastline and limiting the rate of ice flow into the ocean \citep{hooke2005}. Antarctic ice shelves are also important to the formation of Antarctic Bottom Water, a mass of dense water that fills about half of the deep ocean \citep{broecker1998} and that plays an important role in the carbon cycle \citep{marinov2008}.

\begin{figure}
	\centerline{\includegraphics{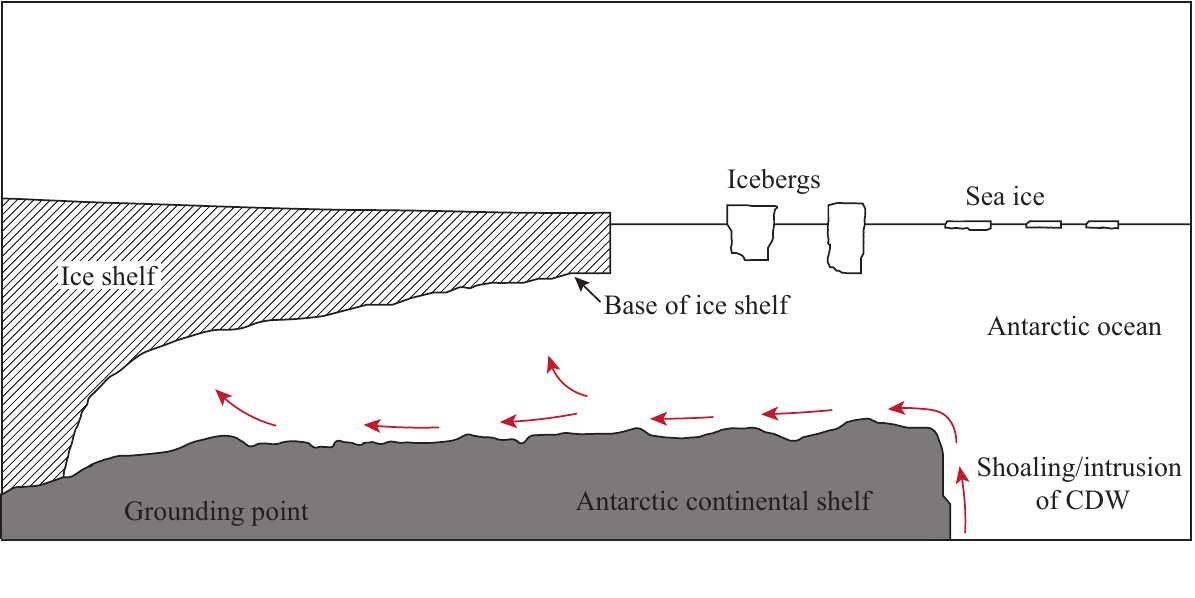}}
	\caption{Warm Circumpolar Deep Water (CDW) rising into the ocean cavity under an Antarctic ice shelf.}
	\label{fig:Antarctica}
\end{figure}

 Recent studies show that warm Circumpolar Deep Water around Antarctica is shoaling  onto the continental shelf and intruding the ocean cavity under ice shelves, causing increased melting of their basal surfaces \citep{jacobs2011, pritchard2012, schmidtko2014}. This process is depicted in figure \ref{fig:Antarctica}. Increased basal melting can trigger the disintegration of ice shelves \citep{feldmann2015collapse} and hence accelerate Antarctic ice loss, which would contribute significantly to global sea level rise. The rough topography of the ocean floor under ice shelves may play a role in guiding the warm shoaling water inside the cavity \citep{brisbourne2014}. Basal melting results in a buoyant plume of meltwater that flows along the shelf base, generating turbulence which in turn affects both the transfer of heat to the shelf and the entrainment of heat from the boundary layer \citep{little2008large}.
 
Previous studies of ice-shelf-ocean interaction have been conducted mainly through numerical models. The heat transfer from the ocean mixed layer to the ice shelf base in these models is parameterized in terms of the temperature difference across and the thermal exchange velocity $\gamma_T$ through the boundary layer at the ice-ocean interface. $\gamma_T$ is defined as the ratio of the thermal diffusivity to the thickness of the boundary layer. In the earlier works of \citet{hellmer1989} and \citet{scheduikat1990}, $\gamma_T$ was taken to be a constant. \citet{jenkins1991} followed the theory of \citet{kader1972}, assuming that the ice--water interface is hydraulically smooth, and expressed $\gamma_T$ in terms of the friction velocity of the turbulent boundary layer. This formulation was used in the studies by \citet{holland2006} and \citet{jenkins2010}. \citet{mcphee1987} developed a parameterization for $\gamma_T$ by using the formulation of \citet{yaglom1974} for the transfer of heat in a turbulent boundary layer near a rough wall and by additionally considering the effect of buoyancy and rotation on heat transfer. \citet{holland1999}, \citet{mueller2012}, and \citet{dansereau2014} adopted this parameterization in their studies. Furthermore, the formation of channels in the ice shelf base as a result of plumes flowing on the underside of the shelf has also been investigated numerically \citep{dallaston2015channelization}.

There are numerous laboratory experiments on heat transfer at a phase change boundary between a solid and a liquid that are relevant to our study. \citet{townsend1964natural} investigated the evolution of the layer of free convection over an ice surface into a stable liquid layer above. The instability of an ice surface, and subsequent formation of a wavy interface, in the presence of a turbulent flow was explored by \citet{gilpin1980wave}. Significant work has been performed on the study of the formation of a mushy layer and on compositional and thermal convection in the liquid during the solidification of a binary solution to explain brine rejection as sea ice forms \citep{huppert1985,wettlaufer1997}. The effect of an external shear flow on a mushy layer has also been investigated \citep{neufeld2008experimental}. In the latter study, a laminar shear flow was applied to an NH$_\textrm{4}$Cl mushy layer from above and the primary focus was the stability of the mushy layer in response to the shear flow. \citet{kerr2015dissolution} developed a theoretical model for the dissolution of a vertical solid surface and tested their model with experimental measurements . These laboratory studies provide an explanation of the physical processes at an ice--water interface and are useful guides for investigating the effect of turbulent warm water at an ice-ocean interface. Also related to ice-shelf-ocean interaction is the set of experiments by \citet{stern2014effect} on the effect of geometry on circulation inside the ice shelf cavity and at the ice shelf front. None of these studies, however, consider the effect of shear-driven turbulence on what is essentially a horizontal ice shelf--ocean interface.

In this paper, we describe an experimental study on the response of an ice--water interface to forced convection in the form of turbulent mixing in water over ice. The experiments are conducted in a tank with a layer of ice growing on a basal cooling plate at one end to represent the base of an ice-shelf and with a rough surface on the other end to represent the rough ocean floor in the far-field. Turbulent mixing causes warm water to be transported from the far-field to the ice--water interface. Our laboratory set-up is thus an idealized model of the ocean cavity under Antarctic ice shelves in which the circulation of relatively warm water is reaching the basal surface of these ice shelves, causing accelerated basal melting.  We formulate a theoretical model for the evolution of the ice thickness in our experiments and compare our measurements with the prediction from our theoretical model in order to develop a parameterization for the turbulent heat transfer at the ice--water interface. The apparatus and procedure are described in \S \ref{sec:expMethod}. In \S \ref{sec:energyBalance}, the governing equations in our theoretical model are outlined. The results from the set of experiments are shown in \S \ref{sec:expResults} and are compared to the theoretical model in \S \ref{sec:modelComp}. In \S \ref{sec:discus}, we discuss the geophysical application of our results. Finally, we summarize our study in \S \ref{sec:summary}.

The dimensionless control parameters that are relevant to the study are the Reynolds number, friction Reynolds number, Rossby number, and Stefan number. The definition of these parameters and their estimated values in our experiments and in an ice shelf cavity are listed in table \ref{tab:dimensionless}. In the definitions, the subscript $s$ refers to the solid (ice) and the subscript $\ell$ refers to the liquid (water). $D$ denotes the depth of the liquid layer; $U_\infty$, the free-stream velocity; $u_\ast$, the friction velocity; $\nu$, the kinematic viscosity; $\Omega$, the angular frequency of rotation; $c$, the specific heat capacity; $\Delta T$, the temperature difference; $L$, the specific latent heat; and $\alpha$, the thermal diffusivity.  

\begin{table}
	\begin{center}
		\def~{\hphantom{0}}
		\begin{tabular}{lccc}
			Non-dimensional number  & Definition &  Experiment & Ice shelf cavity \\[3pt]
			Reynolds, $\Rey$ & $U_\infty D/\nu$ & $10^3$ -- $10^4$  & $10^6$  \\
			Friction Reynolds, $\Rey_*$ & $u_*D/\nu$ & $10^2$ -- $10^3$ &  $10^5$ \\
			Rossby, $Ro$ & $U_\infty/\Omega D$    & 0.7 &  1\\
			Stefan, $St$ & $c_s \Delta T_s/L$ & 0.2 &  0.2 \\
			Prandtl, $\Pran$ & $\nu/\alpha_\ell$ & 13.6 & 13.8 \\ 
			Peclet, $\Pen$ & $U_\infty D/\alpha_\ell$ & $10^4$ -- $10^5$ & $10^7$ \\
			Volumetric heat capacity ratio, $\mathcal{C}$ & $\rho_\ell c_\ell/\rho_s c_s$ & 2.2 & 2.2 \\
			Thermal diffusivity ratio, $A$ & $\alpha_\ell/\alpha_s$ & 0.12 & 0.12
		\end{tabular}
		\caption{Dimensionless control parameters in the experiment and in an ice shelf cavity}
		\label{tab:dimensionless}
	\end{center}
\end{table}

\section{Experimental method} \label{sec:expMethod}
\begin{figure}
	\centerline{\includegraphics{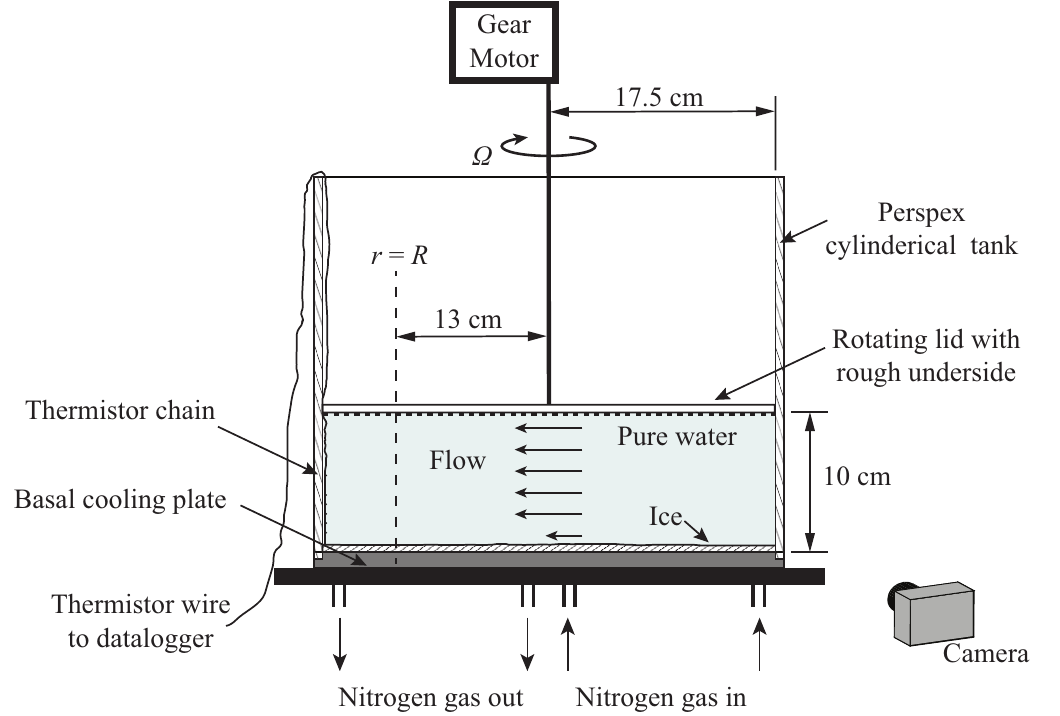}}
	\caption{Schematic diagram of the apparatus}
	\label{fig:Apparatus}
\end{figure}

The experimental apparatus is shown in figure \ref{fig:Apparatus}. It consists of a cylindrical tank of diameter 35 cm with 1.5-cm-thick Perspex walls and a 5-cm-thick aluminum basal cooling plate. The tank is filled with pure water to a height of 10 cm, and  ice is grown by circulating cold nitrogen gas inside the basal cooling plate.  The physical properties of liquid water and ice are listed in table \ref{tab:properties}. The interior of the plate consists of two sets of parallel spiral grooves, one set having an inlet at the center and an outlet near the rim and the other set having an inlet near the rim and an outlet at the center. This arrangement helps achieve a uniform heat flux through the plate and hence uniform ice growth on its surface. The nitrogen flow rate is held constant at 0.14 m$^3$min$^{-1}$ within and across experiments. A perspex cover lid is positioned at the upper surface of the water layer, connected to a gear motor by means of a vertical metal rod. A plastic grid is attached to the underside of the cover lid, creating a rough surface for generating the turbulent shear flow. The grid consists of a lattice of squares, each square having sides of length 1.4 cm and projecting downward beneath the lid a distance of 0.9 cm. The rotation of the cover lid and plastic grid is controlled by the gear motor.

To start each experiment, the water layer, initially at rest and at room temperature, is suddenly cooled from below by turning on the flow of nitrogen into the basal cooling plate. It typically takes about 30 minutes for ice to begin to nucleate on the basal plate. The ice is allowed to grow for another 30 minutes, reaching a nearly uniform thickness of 8-12 mm, depending on the initial temperature of the water. The motor is then turned on, rotating the lid and grid at a constant angular velocity, typically for about one hour. We experimented with lid angular velocities between 0.27 and 1.43 rads$^{-1}$, fast enough to generate a turbulent shear flow in each case. 

Pictures of the ice are taken from the side of the tank at 1-minute intervals with a Nikon D800 camera. The ice thickness is subsequently measured from these pictures using GraphClick, a digitizer software. Seven thermistors are placed on a 5.25-cm-long vertical strip starting from the bottom of the tank to measure temperature at the locations shown in figure \ref{fig:thermistors}. The strip is placed along the wall of the tank and the thermistors protrude 1 cm into the tank. The thermistors are connected to a datalogger. We focus on the ice thickness at a radial distance $R=13$ cm from the tank center, that is, 4.5 cm from the outer wall. This location is a compromise between its proximity to the thermistor chain and its separation from the immediate effects of the outer wall.

Both the friction velocity and fluid velocity of the turbulent shear flow are measured over the entire range of lid angular velocities.The average shear stress, and hence friction velocity, is obtained by measuring the torque on the lid with a torque meter. The fluid velocity is obtained from planar PIV measurements. The water is seeded with nearly spherical glass beads of specific gravity 1.1 and average diameter 10 $\mu$m and illuminated with a pulsed Nd:YAG laser sheet. A vertical light sheet is set up along a chord at $r_i$ to measure the vertical profile of the azimuthal component of the velocity. To measure the radial component of the velocity, a horizontal light sheet is set up to illuminate a sector of the tank's circular cross-sectional area at different heights above the bottom plate. A high-speed CMOS camera synchronized with the pulsed laser taking double exposure images at a resolution of 2048 $\times$ 2048 pixels is positioned in the looking downward and looking sideward orientations, for imaging the radial flow and azimuthal flow, respectively. The open source software PIVlab \citep{thielicke2014pivlab} is used to calculate PIV velocities from the exposures. For the determinations of the azimuthal velocity profiles, only a vertical strip at the center of the sideward looking images, where the particles move in the plane of the light sheet, is used in the analysis.

\section{Ice energy balance}\label{sec:energyBalance}
\begin{figure}
	\centerline{\includegraphics{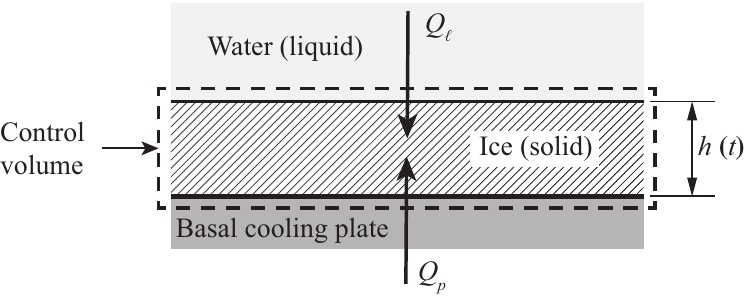}}
	\caption{Control volume around ice}
	\label{fig:ControlVolume}
\end{figure}

The energy (enthalpy) balance in a control volume enclosing the ice with thickness $h$ at a time $t$ shown in figure \ref{fig:ControlVolume} yields the following relationship:
\begin{equation}
\frac{\D E}{\D t} = Q_p + Q_\ell,
\label{hbal}
\end{equation}
where $E$ is the energy (enthalpy) content of the ice, the subscript $p$ refers to the plate, and $Q$ is the heat entering the control volume from the region denoted by its subscript. Considering a one-dimensional energy balance, $E$ and $Q_p$ can be expressed as
\begin{equation}
E = \rho_s c_s \int_0^h T_s \ \D z - \rho_s L h
\label{ice_enthalpy}
\end{equation}
and
\begin{equation}
Q_p = - \frac{k_s \Delta T_s}{h}.
\label{qp}
\end{equation}

In (\ref{ice_enthalpy}) and (\ref{qp}), $k$ is thermal conductivity, $T$ is temperature, and $\Delta T_s = T_f - T_p$, where $T_f$ is the freezing temperature of water (also the temperature of the ice--water interface). Assuming that the temperature varies linearly in the vertical direction through the ice, $T_s=(T_f + T_p)/2$. Numerical values of the physical properties of water and ice are given in table \ref{tab:properties}. The first term on the right-hand side in (\ref{ice_enthalpy}) can be rewritten as
\begin{equation}
\rho_s c_s \int_0^h T_s \ \D z = \rho_s c_s h \bigg (\frac{\Delta T_s}{2} + T_p \bigg ).
\label{ice_enthalpy2}
\end{equation}

\subsection{No turbulent mixing }
When the ice is growing in quiescent water, heat is transferred by conduction. We ignore free convection in the liquid. In this case $Q_\ell$ is the sum of the conductive heat transfer through and the rate of change of enthalpy of the liquid, so that
\begin{equation}
Q_\ell =  \frac{k_\ell \Delta T_\ell}{\delta} + \rho_s c_s T_f \frac{\D h}{\D t}
\label{ql}
\end{equation}
where the subscript $\ell$ refers to the liquid properties, $\delta$ is the thickness of the thermal boundary layer above the ice, and $\Delta T_\ell = T_\infty - T_f$, $T_\infty$ being the temperature of the liquid far-field. Introducing
\begin{equation}
\delta = \frac{\alpha_\ell}{\D h/\D t}
\end{equation}
from the heat balance in a control volume in the liquid region above the ice--water interface, with $\alpha_\ell$ given by
\begin{equation}
\alpha_\ell = \frac{k_\ell}{\rho_\ell c_\ell},
\end{equation}
the conductive term in (\ref{ql}) can be rewritten as
\begin{equation}
\frac{k_\ell \Delta T_\ell}{\delta} = \rho_\ell c_\ell \Delta T_\ell \frac{\D h}{\D t}.
\label{liquid_conductive}
\end{equation}
Substitution of (\ref{ice_enthalpy}), (\ref{qp}), (\ref{ice_enthalpy2}), (\ref{ql}), and (\ref{liquid_conductive}) into (\ref{hbal}) gives, for the heat balance in the control volume,
\begin{equation}
\bigg ( \rho_s L + \frac{\rho_s c_s \Delta T_s}{2} + \rho_\ell c_\ell \Delta T_\ell \bigg ) \frac{\D h}{\D t} = \frac{k_s \Delta T_s}{h} + \frac{\rho_s c_s h}{2} \frac{\D}{\D t} (\Delta T_s),
\end{equation}
which can be approximated, based on the largest terms, as
\begin{equation}
( \rho_s L + \rho_\ell c_\ell \Delta T_\ell ) \frac{\D h}{\D t} \simeq \frac{k_s \Delta T_s}{h}.
\label{hbal_p1}
\end{equation}
This equation is non-dimensionalized by taking the length scale to be the depth $D$ of the liquid, the temperature difference scale to be the temperature difference across the solid at the onset of turbulent mixing, and the time scale to be $D^2/\alpha_\ell$, which corresponds to the characteristic time for thermal diffusion over the distance $D$. This yields 
\begin{equation}
\bigg [ \frac{1}{St} + \mathcal{C}^\ast (\Delta T_\ell)^\ast \bigg ] \frac{\D h^\ast}{\D t^\ast} = \frac{1}{A} \frac{(\Delta T_s)^\ast}{h^\ast}.
\label{dimensionless_p1}
\end{equation}
where variables with a superscript $\ast$ are in non-dimensional form. $\mathcal{C} = \rho_\ell c_\ell/\rho_s c_s$, the ratio of the volumetric heat capacity of the liquid to that of the solid, and $A = \alpha_\ell/\alpha_s$, the ratio of the thermal diffusivity of the solid to that of the liquid. The typical values of $\mathcal{C}$ and $A$ for the laboratory experiment and for the geophysical application are listed in table \ref{tab:dimensionless}.

\subsection{Turbulent mixing}
For turbulent flow over a flat plate at constant temperature, Reynolds analogy relates the convective heat flux $q_T$ to the properties of the momentum boundary layer. In Reynolds analogy, the heat flux and momentum flux at the plate in a turbulent boundary layer are considered equivalent since they are both influenced by the turbulent motion above the plate. The expression for $q_T$ \citep[see][p. 564]{white1974} is
\begin{equation}
q_T = \rho_\ell U_\infty c_\ell \Delta T_\ell C_h
\label{qt}
\end{equation}
where $C_h$ is a heat-transfer coefficient (Stanton number) given empirically by
\begin{equation}
	C_h = \frac{c_f/2}{1 + 12.8 (\Pran^{0.68} -1 )\sqrt{c_f/2}}. 
\end{equation}
$U_\infty$ is the velocity of the liquid in the far-field, $\Pran$ is the Prandtl number, and $c_f$ is the coefficient of friction defined as
\begin{equation}
c_f = 2 \frac{u_\ast^2}{U_\infty^2}
\label{cf}
\end{equation}
where $u_*$ is the friction velocity. We introduce the coefficient G in the expression for $C_h$ to substitute for the constant term $12.8(\Pran^{0.68} -1)$. In the context of our experiment, this term is the term we are trying to constrain. For the turbulent mixing phase in our experiments, $Q_\ell$ is augmented by $q_T$, and hence the energy balance for the control volume becomes
\begin{equation}
( \rho_s L + \rho_\ell c_\ell \Delta T_\ell ) \frac{\D h}{\D t} \simeq \frac{k_s \Delta T_s}{h} - \rho_\ell U_\infty c_\ell \Delta T_\ell C_h.
\label{hbal_p2}
\end{equation}
By using the same length, temperature difference, and time scales as in (\ref{dimensionless_p1}) and by additionally using $U_\infty$ as the velocity scale, this expression is non-dimensionalized to obtain
\begin{equation}
\bigg [ \frac{1}{St} + \mathcal{C} (\Delta T_\ell)^\ast \bigg ] \frac{\D h^\ast}{\D t^\ast} = \frac{1}{A} \frac{(\Delta T_s)^\ast}{h^\ast} - \Rey \Pran \mathcal{C} (\Delta T_\ell)^\ast C_h,
\label{dimensionless_p2}
\end{equation}
where $\Rey \Pran = \Pen$. Table \ref{tab:dimensionless} lists typical values of $\Pen$ for the sub-ice shelf cavity and the laboratory set-up.

\begin{table}
	\begin{center}
		\def~{\hphantom{0}}
		\begin{tabular}{lccc}
			Property  & Units   &   Liquid water & Solid ice \\[3pt]
			Freezing temperature, $T_f$   & K &  & 273.15 $^b$ \\
			Density, $\rho$   & kg m$^{-3}$ & 999.8 $^a$ & 916.7 $^b$ \\
			Specific latent heat of fusion, $L$   & J kg$^{-1}$ &  & 3.33 $\times 10^{5}$ $^b$ \\
			Isobaric specific heat capacity, $c$  & J kg$^{-1}$ K$^{-1}$ & 4.21 $\times 10^{3} \ $$^a$ & 2.10 $\times 10^{3} $ $^b$ \\
			Thermal conductivity, $k$   & W m$^{-1}$ K$^{-1}$ & 0.556 $^b$ & 2.16 $^b$ \\
			Kinematic viscosity, $\mu$ & Pa s & 1.79 $\times 10^{-3} $ $^b$ & \\
			Thermal expansion coefficient, $\beta$ & K$^{-1}$ & 6.77 $\times 10^{-5}$ $^c$ & 160 $^b$
		\end{tabular}
		\caption{Physical properties of liquid water and ice at standard atmospheric pressure. $\rho$, $L$, $c$, $k$, $\mu$, and $\beta$ at 273.15 K. $\beta$ for liquid water is linear and $\beta$ for solid ice is volumetric. $^a$ From \citet{wagner2002}, $^b$ from \citet{haynes2015crc}, and $^c$ from \citet{iociapso}.}
		\label{tab:properties}
	\end{center}
\end{table}

\section{Experimental results}\label{sec:expResults}
We conducted a set of eleven experiments at different angular velocities of rotation of the lid $\Omega$. The value of $\Omega$ for each experiment is listed in table \ref{tab:expList}. Experiment 0 is a null experiment in which the lid was not rotated, and hence the water was not mixed by turbulence over the whole duration. The lid $\Rey$ at a radius $r$, $\Rey_r$, in the tank is defined as
\begin{equation}
\Rey_r = \frac{(\Omega r) D}{\nu}.
\label{reynolds}
\end{equation}
When $\Omega r=U_\infty$, this definition of $\Rey$ is the same as in table \ref{tab:dimensionless}. The value of $\Rey_r$ at $r=R$, which we denote by $Re_R$, for each experiment is given in table \ref{tab:expList}.  We refer to the first portion of each experiment in which ice grows by conduction in still water as Phase 1 and the second portion of each experiment in which there is a turbulent shear flow and mixing as Phase 2.

\subsection{Ice thickness}
\begin{figure}
	\centerline{\includegraphics{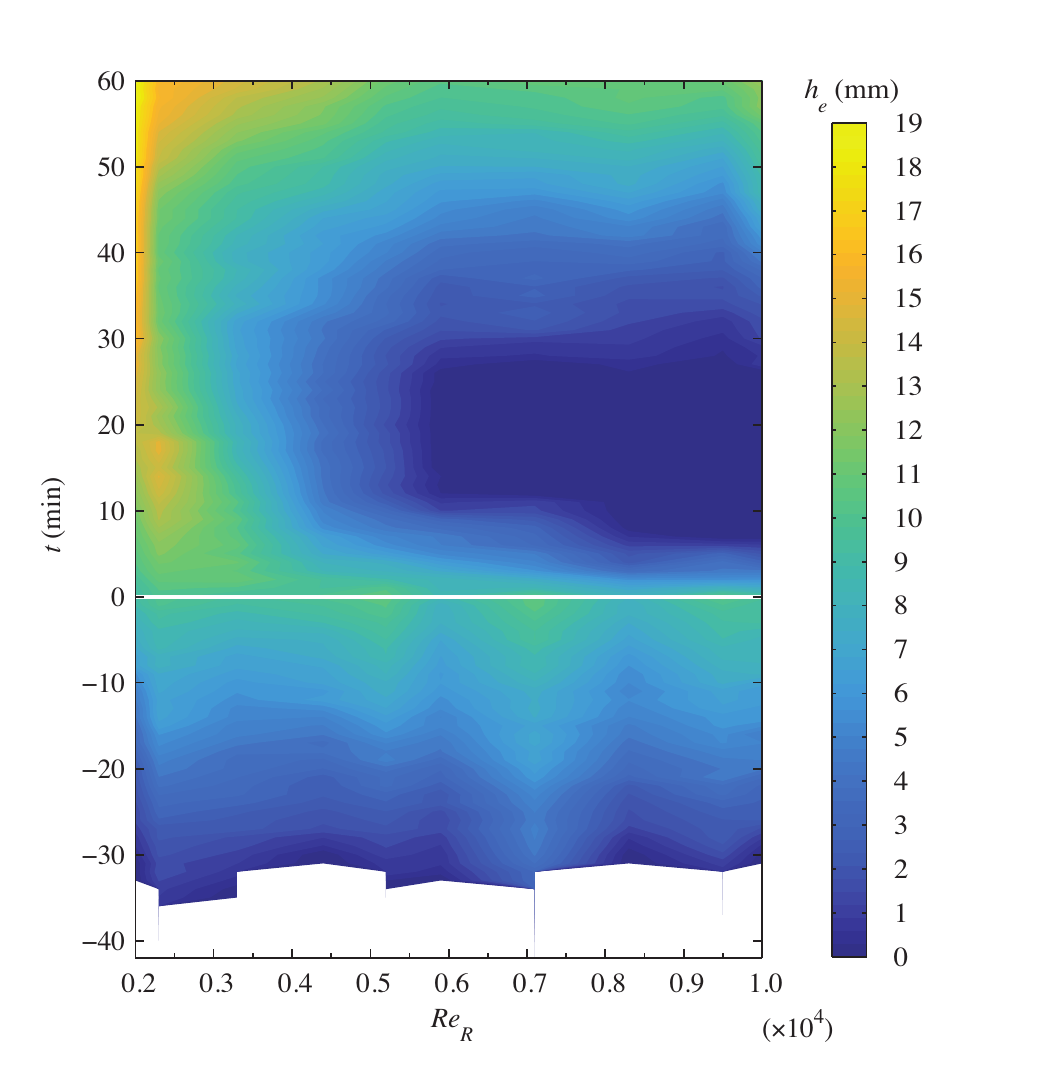}}
	\caption{Ice thickness at $R$ over the course of Experiments 1-10. The horizontal axis denotes the range of $\Rey_R$ covered by the experiments and the vertical axis denotes the time relative to the onset of mixing in each experiment. The horizontal white line indicates the onset of mixing. }
	\label{fig:ContourThickness}
\end{figure}

\begin{figure}
	\centerline{\includegraphics{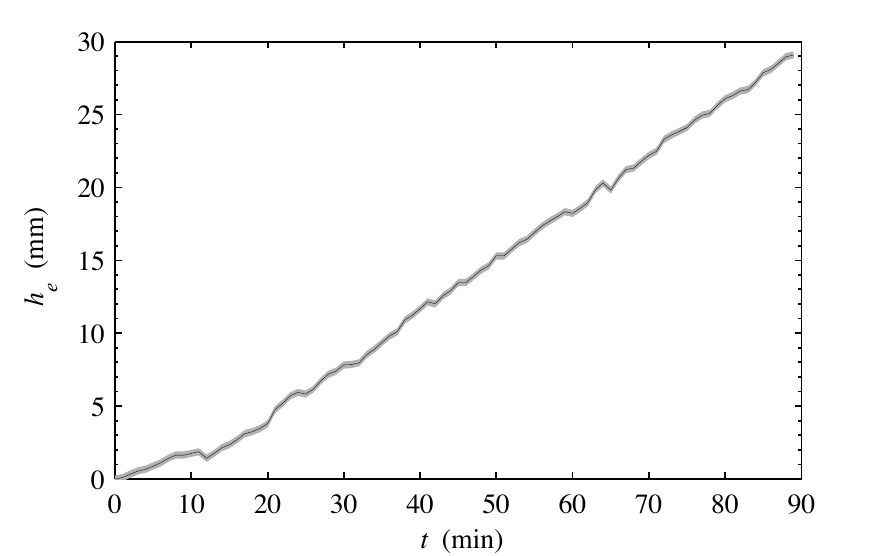}}
	\caption{Ice growth in Experiment 0 (with no shear flow)}
	\label{fig:NullExperiment}
\end{figure}

\begin{table}
	\begin{center}
		\def~{\hphantom{0}}
		\begin{tabular}{l c c c}
			Experiment & $\Omega$ (rad/s) & Lid $\Rey_R$ & Regime\\[3pt]
			0 & 0 & 0 & \\
			1 & 0.27 & 2.0$\times 10^3$ & Attenuated growth\\
			2 & 0.32 & 2.3$\times 10^3$ & Partial Melting \\
			3 & 0.45 & 3.3$\times 10^3$ & Partial Melting \\
			4 & 0.60 & 4.4$\times 10^3$ & Partial Melting \\
			5 & 0.71 & 5.2$\times 10^3$ & Partial Melting \\
			6 & 0.82 & 5.9$\times 10^3$ & Complete Melting \\
			7 & 0.98 & 7.1$\times 10^3$ & Complete Melting\\
			8 & 1.14 & 8.3$\times 10^3$ & Complete Melting\\
			9 & 1.31 & 9.5$\times 10^3$ & Complete Melting\\
			10 & 1.43 & 1.0$\times 10^4$ & Complete Melting\\
		\end{tabular}
		\caption{Angular frequency $\Omega$ of lid and lid $\Rey_R$ in experiments}
		\label{tab:expList}
	\end{center}
\end{table}
Measured ice thickness $h_e$ versus time are shown in figures \ref{fig:ContourThickness} and \ref{fig:NullExperiment}, from Experiments 1--10 and Experiment 0 respectively. In figure \ref{fig:ContourThickness} the time $t=0$ corresponds to the onset of the turbulent shear flow. In figure \ref{fig:NullExperiment}, the shaded region along the line plot has a total width of 0.5 mm and represents the error in the measurements. The error was estimated by taking the standard deviation of 10 repeated measurements of the ice thickness at $R$ during Phase 1 of a typical experiment. The ice thickness measurements in Experiment 0 and in Phase 1 of Experiments 1--10 are assigned the same error estimate.

The error in Phase 2 measured by the same method, using Phase 2 measurements from a typical experiment, is 0.9 mm. The error in ice thickness measurements in Phase 2 is larger than in Phase 1 because the ice--water interface becomes wavy when ice melts in the presence of the turbulent flow. It is difficult to visually identify the ice thickness along the diameter from the side-view pictures of the tank due to the waviness of the interface.

Ice grows at an almost constant rate when the water is undisturbed, as in Experiment 0 and in Phase 1 of Experiments 1--10. In Phase 2, mixing by the turbulent shear flow transports warm water from the far-field to the ice--water interface, which promotes heat transfer to the ice. The ice then responds in one of three ways, each of which we have observed as a transient at our measurement location $R$: (1) attenuated ice growth, (2) partial melting, and (3) complete melting. Following this transient response, re-growth of ice at the same rate as in Phase 1 is observed.

\begin{figure}
	\centering
	\begin{subfigure}
		\centering
		\includegraphics{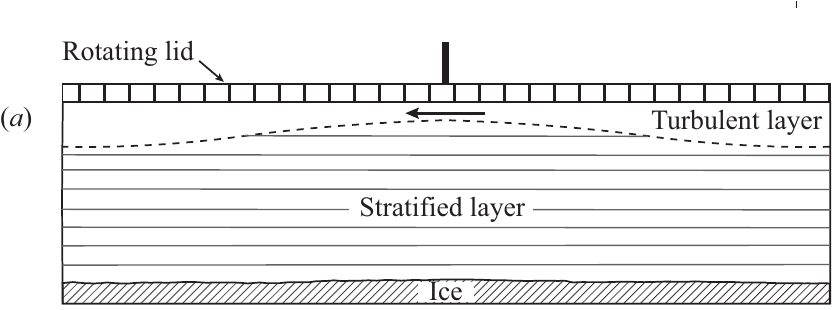}
	\end{subfigure}
	\begin{subfigure}
		\centering
		\includegraphics{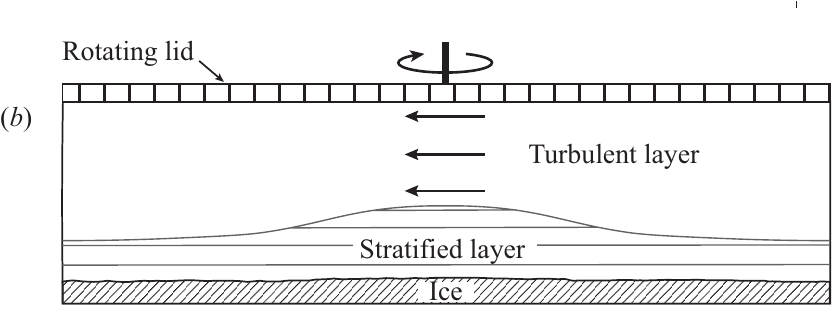}
	\end{subfigure}
	\begin{subfigure}
		\centering
		\includegraphics{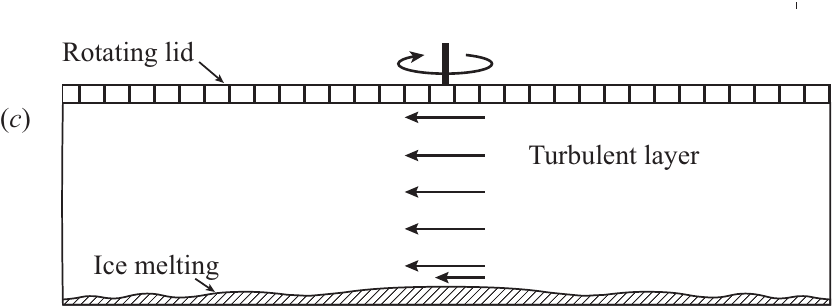}
	\end{subfigure}
	\caption{Changes in the liquid layer following the onset of turbulent mixing. At the lower end of experimental $\Rey_R$, the steps shown in this diagram occur over several minutes, whereas at the upper end of experimental $\Rey_R$, they occur in a few seconds. (\textit{a}) A stratified layer initially separates the turbulent layer from the ice surface. The interface between the stratified layer and turbulent layer is dome-shaped. (\textit{b}) The turbulent layer progresses downwards, eroding the stratified layer. (\textit{c}) The turbulent layer has reached the ice--water interface and causes the ice to melt. The thickness of ice melted increases with radius. A spiral wavy profile develops on the ice surface during melting. }
	\label{MixingOnset}
\end{figure}

Figure \ref{MixingOnset} shows the sequence of structures that are observed in the ice--water system following the onset of turbulent mixing. A thermally stratified water layer initially separates the growing ice from the turbulent flow, as the turbulence develops beneath the rotating lid. The interface between the stratified layer and the turbulent flow is dome-shaped because the turbulent shear stress $\tau$ increases proportionally to $r^2$ and is therefore weaker near the center of the tank. In our lowest $\Rey_R$ experiment, the stratified layer persists in the presence of the shear flow, thereby preventing turbulence from reaching the ice--water interface. Ice growth is attenuated in this case, but not stopped. At the other extreme, in our highest $\Rey_R$ experiments, the turbulent mixing is strong enough to erode the stratified layer entirely almost immediately after the onset of the turbulent shear flow. When the turbulence comes in direct contact with the ice--water interface, it produces complete melting at high $\Rey_R$ and partial melting at intermediate $\Rey_R$. In the partial melting cases, the thickness of ice melted increases with radial distance from the tank center. In table \ref{tab:expList}, the transient behavior the ice adopts at $R$ in response to turbulent mixing in Experiments 1--10 is given along with the corresponding $\Rey_R$.

When ice melts in our experiments, a spiral ripple pattern develops on the ice--water interface, as mentioned previously. This phemomenon has been explained by \citet{gilpin1980wave}, as follows. Although the ice thickness is approximately uniform at the end of Phase 1, there are nevertheless small-amplitude deviations from uniform thickness due to random perturbations and minor design flaws in the cooling apparatus. \citet{gilpin1980wave} found that such an interface will be unstable to growth in the presence of turbulence when the heat flux from the liquid to the solid is large, which is the case in our experiments during transient melting. The mechanism for the instability involves flow separation downstream of an irregularity in the ice, which causes the heat transfer at a crest to be smaller than the
heat transfer at a valley. The amplitude of the irregularity thus grows, which further amplifies the irregularity in the shear flow, producing a growing set of undulations on the ice--water interface as it melts. The wavelength of the undulations increases with $u_\ast$, which is proportional to $r$ in our experiments. The dependence of the undulation wavelength on distance from the tank center gives rise to the spiral profile of the ice--water interface undulations that we observe.

\subsection{Temperature}
\begin{figure}
	\centerline{\includegraphics{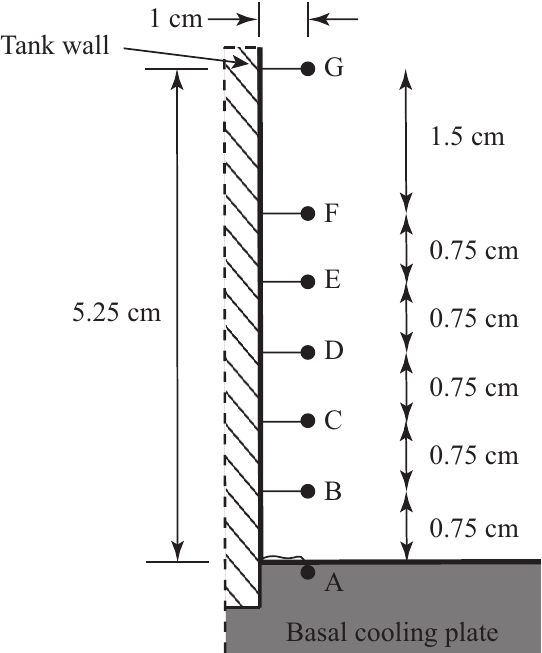}}
	\caption{Arrangement of thermistors}
	\label{fig:thermistors}
\end{figure}

Thermistors A--G are used to measure the temperature at the heights indicated in figure \ref{fig:thermistors}. The resistance $R$ of a thermistor is related to its temperature $T$ according to the Steinhart-Hart equation,  
\begin{equation}
\frac{1}{T} = a_1 + a_2 \ln R + a_3 (\ln R)^3,
\label{SteinhartHart}
\end{equation}
where $a_1$, $a_2$, and $a_3$ are the Steinhart-Hart coefficients and are unique to each thermistor. We obtained these coefficients by calibration prior to our series of experiments.

\begin{figure}
	\centerline{\includegraphics{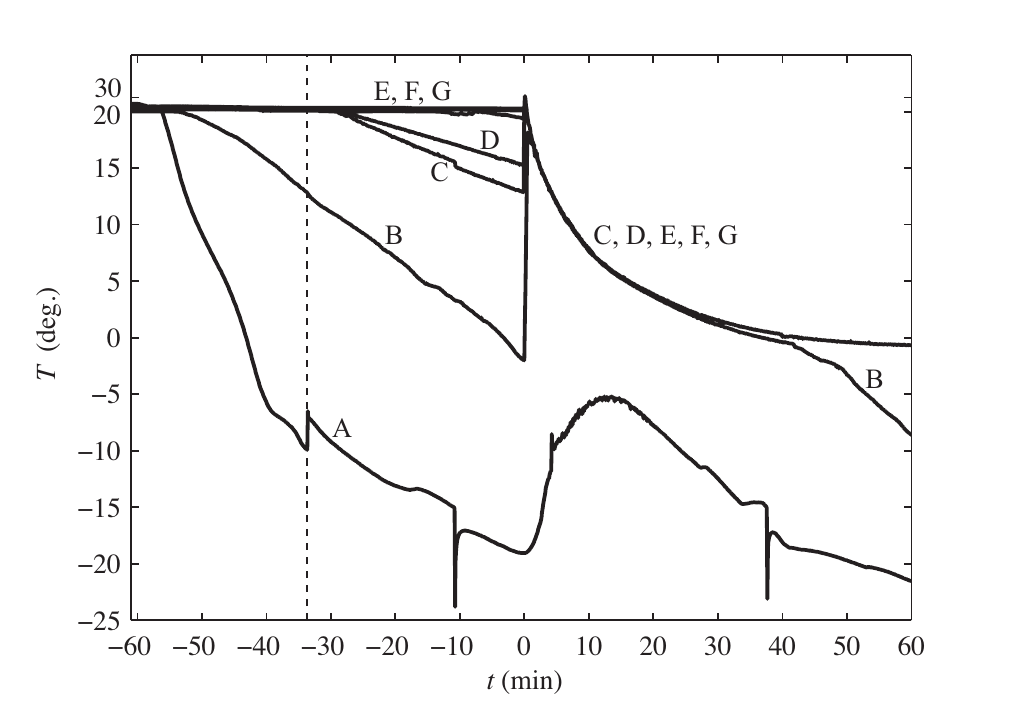}}
	\caption{ Temperature recorded by thermistors A, B, C, D, E, F, and G over the course of a typical experiment. The horizontal axis denotes time relative to the onset of turbulent mixing. The vertical dashed line indicates the time at which ice forms a thin layer on the bottom plate.}
	\label{fig:TemperatureTime}
\end{figure}

Temperature is recorded starting from the instant nitrogen starts to circulate inside the basal cooling plate. The evolution of temperature at the thermistor locations A--G in a typical experiment (in this case, Experiment 6) is shown in figure \ref{fig:TemperatureTime}. Following the onset of turbulent mixing, there is an increase in temperature at A because warm water transported to the bottom of the tank causes melting of ice. At that time, thermistors B, C, D, E, F, and G record the same temperature, signaling that mixing results in a homogeneous distribution of temperature in the turbulent shear flow. Note that after about 40 minutes of cooling, the temperature at B departs from the temperatures at C, D, E, F, and G as thermistor B is engulfed by the growing ice. 

\begin{figure}
	\centering
	\begin{subfigure}
		\centering
		\includegraphics{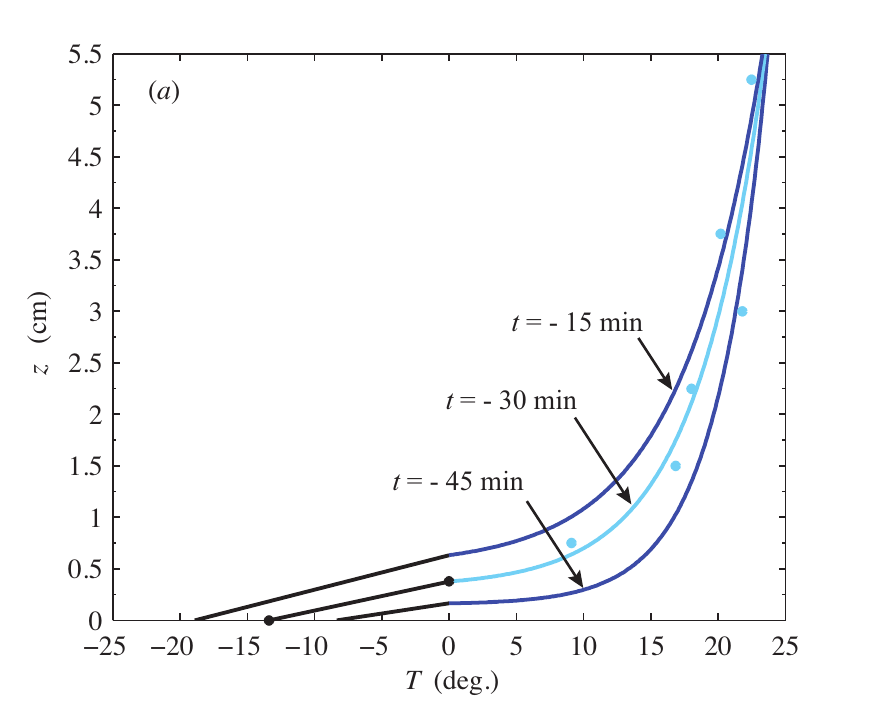}
	\end{subfigure}
\quad	
	\begin{subfigure}
		\centering
		\includegraphics{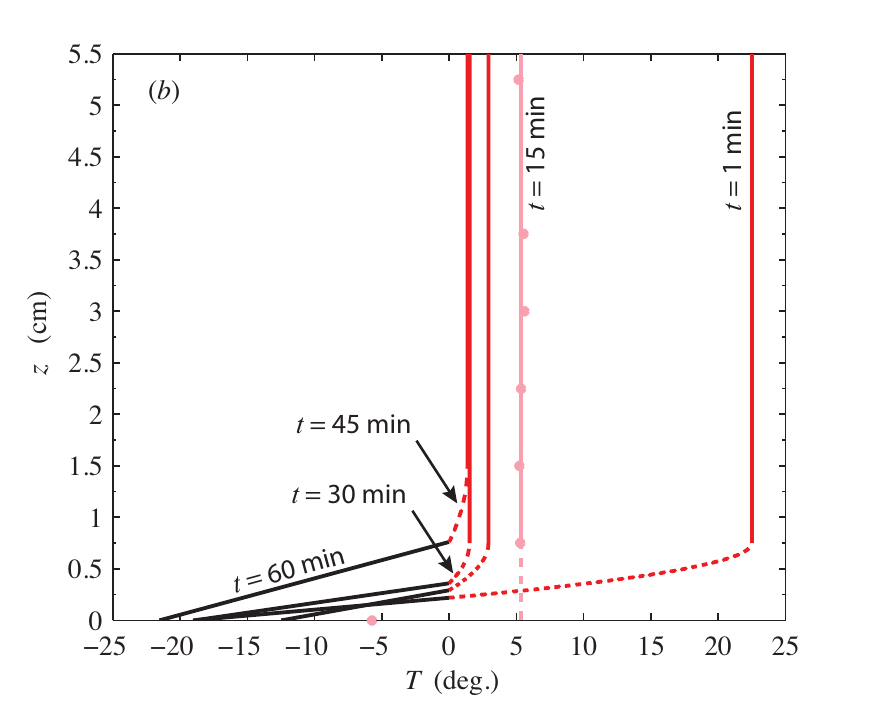}
	\end{subfigure}
	\caption{Vertical profiles of temperature at different times relative to the onset of turbulent mixing during (\textit{a}) Phase 1 and (\textit{b}) Phase 2 of a typical experiment. The temperature data points are shown for the $t=-30$ min profile in (\textit{a}) and for the $t=15$ min profile in (\textit{b}). In (\textit{b}), a dashed line is drawn between the location of the ice water interface and the location of the first thermistor above the ice--water interface to indicate a possible temperature profile in that layer. (Colour online) Liquid temperature profiles in Phase 1 are in blue and liquid temperature profiles in Phase 2 are in red.}
	\label{fig:TemperatureHeight}
\end{figure}

Vertical profiles of temperature in the same experiment at the indicated times during Phase 1 and Phase 2 are shown in figures \ref{fig:TemperatureHeight}(\textit{a}) and \ref{fig:TemperatureHeight}(\textit{b}) respectively. Because the ice--water interface is at $T_f$ = 0 \textdegree C, the measured ice thickness at any time can be checked by interpolating $T_f$ in the temperature time series recorded by thermistors A--G. The temperature within the ice can safely be assumed to increase linearly from the temperature of the plate to $T_f$, because heat transfer in the ice is by conduction and its growth rate is slow enough that the ice is in thermal equilibrium with its boundaries.  The linear distributions of ice temperature are shown by the straight lines in the left half of figures \ref{fig:TemperatureHeight}(\textit{a}) and \ref{fig:TemperatureHeight}(\textit{b}). 
Temperature profiles above the ice--water interface in Phase 1 (figure \ref{fig:TemperatureHeight}\textit{a}) are exponential fits to the temperature measurements. For clarity, the temperature data points corresponding to only one profile have been included in each figure. The fitted liquid layer temperature profiles in Phase 1 are characteristic of heat transfer in the liquid by conduction only. 

We saw no evidence of natural convection in the liquid layer during Phase 1. For a liquid water layer over an ice--water interface at 0 \textdegree C, natural convection onsets at Rayleigh numbers above 1700, which has been confirmed experimentally by \cite{boger1967effect}. The Rayleigh number for this system can be expressed as
\begin{equation}
\textrm{Rayleigh} = \frac{d^3 \beta \rho_\ell^2 g c_\ell (\Delta T) }{\mu k_\ell}
\label{Ra}
\end{equation}
where $\beta$ is the thermal expansion of water, $d$ is the convecting layer depth, and $\Delta T$ is the temperature difference across $d$. \cite{boger1967effect} take $d$ to be the thickness of the liquid layer between the ice--water interface and the height at which the water is at 4 \textdegree C, where it has maximum density. Interpolating the value of $d$ from the temperature measurements in Phase 1 of our experiments and using values from table \ref{tab:properties} for the physical properties of water, we found that the Rayleigh number in Phase 1 varies from 250 to about 850. It therefore remains below the critical Rayleigh number at which natural convection would occur. 

During Phase 2 of these experiments, all of the thermistors in the liquid typically record nearly the same temperature at a given time after the initial thermal stratification in the liquid has been destroyed by turbulent mixing. A vertical line through the mean of the temperatures measured by the thermistors in the liquid is drawn in figure \ref{fig:TemperatureHeight}(\textit{b}) to represent a homogeneous vertical temperature profile in the liquid layer.

\begin{figure}
	\centerline{\includegraphics{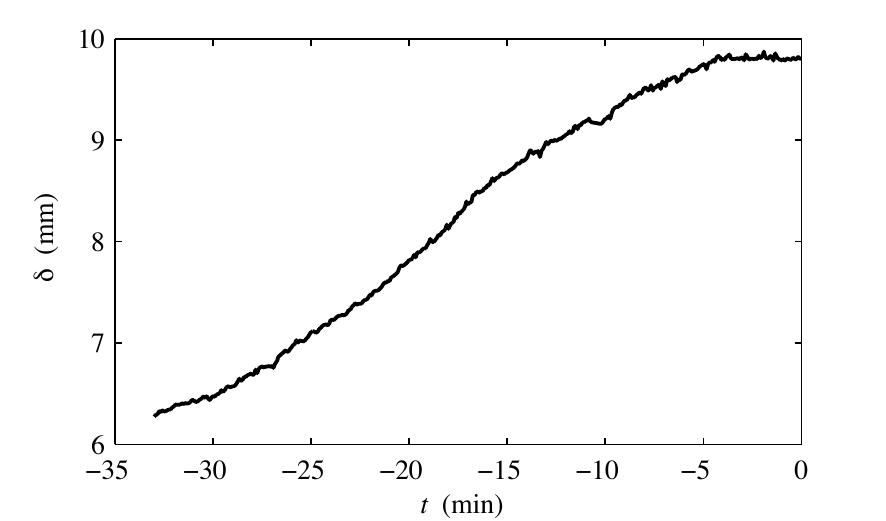}}
	\caption{Boundary layer thickness $\delta$ calculated from exponential fit through non-dimensionalized vertical temperature time series}
	\label{fig:BL}
\end{figure}

As the liquid cools by conduction in Phase 1 of the experiment, the temperature $T$ in its thermal boundary layer can be modelled according to the relation
\begin{equation}
\frac{T - T_f}{T_\infty - T_f} = 1-{\rm{e}}^{-(z-h_e)/\delta}, \ z \ge h_e.   
\label{tbl}
\end{equation}
The thickness $\delta$ of the thermal boundary layer can be obtained by taking the reciprocal of the fit coefficient of an exponential fit to this relationship. Figure \ref{fig:BL} shows the evolution of $\delta$ calculated in this way in Phase 1 of the same experiment. The initial value of $\delta$ is non-zero because prior to the formation of ice, a thermal boundary layer was already present in the liquid due to cooling from the bottom by the basal cooling plate. The thickness of the thermal boundary layer increases from its initial value as Phase 1 of the experiment proceeds. This indicates that, for a control volume in the liquid above the ice--water interface, the heat loss by conduction to the ice is larger than the enthalpy decrease of the control volume due to the movement of the ice--water interface into it. At the end of Phase 1, $\delta$ asymptotes to a uniform value. At this stage, the energy balance in the control volume above the ice--water interface is at steady-state, that is, conductive heat loss to the ice is balanced by enthalpy decrease due to the upward movement of the ice--water interface. On the basis of a simple one-dimensional model for the energy balance in a control volume in which the motion of the liquid is due to the motion of the bottom boundary, $\delta = \alpha_\ell/(\D h_e/\D t)$. The predicted value of $\delta$ from this model for our case is 20 mm, which is about twice larger than the steady-state value of $\delta$ from figure \ref{fig:BL}. The rate of growth of ice, $\D h_e/\D t$, is very small in our experiment (about $5.5 \times 10^{-3} \  \textrm{mms}^{-1}$). There is very little liquid convective motion in the control volume above the ice--water interface in response to the very-slowly-moving interface, and hence, the thermal boundary layer is thinner than the theoretical prediction. 

\subsection{Velocity}
Application of the heat balance shown in (\ref{hbal_p2}) to the control volume around the ice--water interface in figure \ref{fig:ControlVolume} requires knowledge of the fluid velocity in the far-field. Since the temperature distribution in the liquid is nearly homogeneous when there is turbulent mixing, buoyancy forces in the liquid are weak during this phase of the experiments. The circulation in the far-field is thus due to the shear induced by the rotating lid only. 

\begin{figure}
	\centerline{\includegraphics{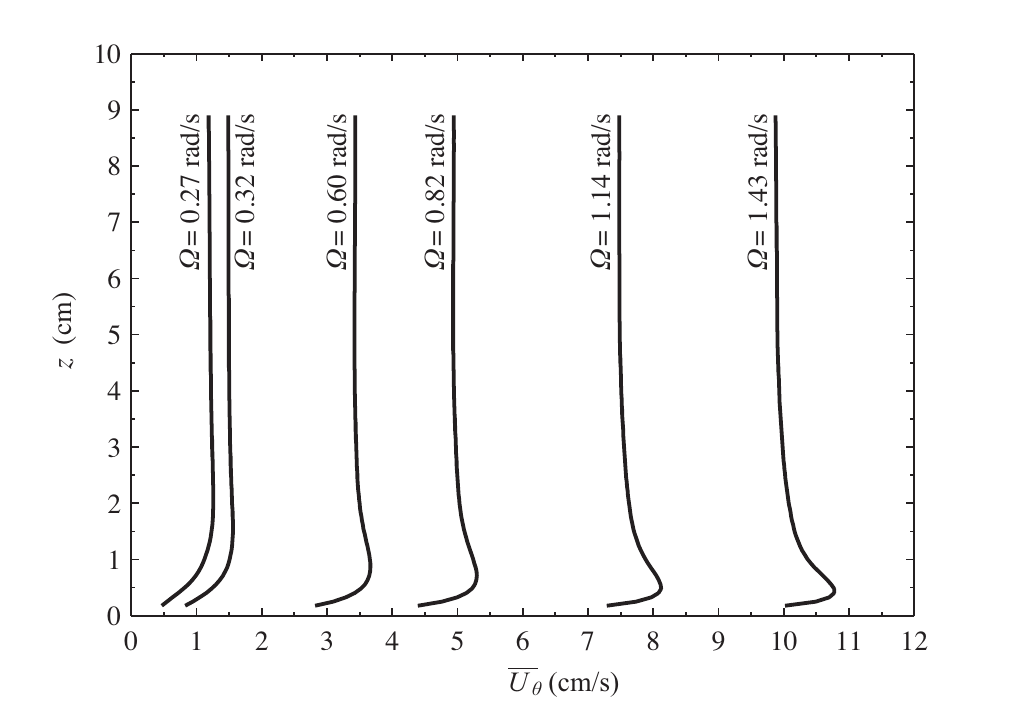}}
	\caption{Mean azimuthal velocity in the fluid column at $r_i$ for different angular velocities of the lid}
	\label{fig:Vtheta}
\end{figure}

\begin{figure}
	\centering
	\begin{subfigure}
		\centering
		\includegraphics{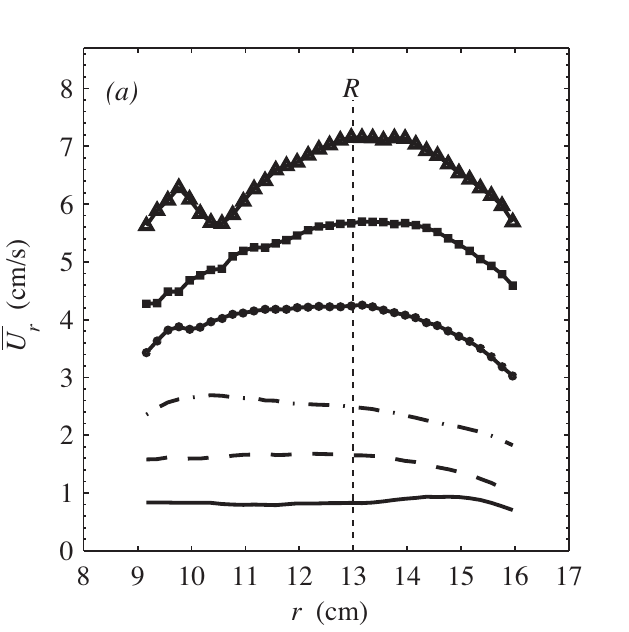}
	\end{subfigure}
	\quad
	\begin{subfigure}
		\centering
		\includegraphics{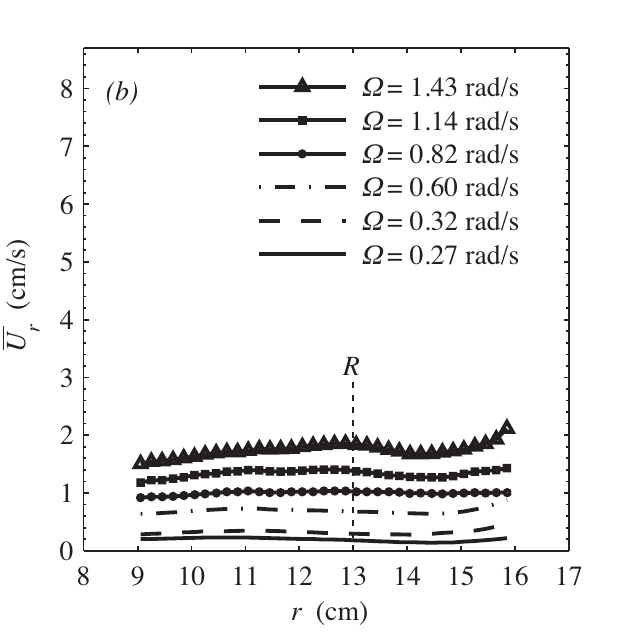}

	\end{subfigure}
	\caption{Mean radial velocity at heights of (\textit{a}) 0.5 cm and (\textit{b}) 7 cm above the bottom plate. Positive values correspond to inward direction. The vertical dashed line in (\textit{a}) and (\textit{b}) is at $r_i$ where the ice measurements are taken.}
	\label{fig:Vr}
\end{figure}

The velocity of the shear-driven turbulent flow above the flat bottom surface of the tank is measured for the purpose of relating the fluid velocity in the far-field to the lid velocity. We denote by $\overline{U_\theta}$ the mean of the azimuthal velocity component and by $\overline{U_r}$ the mean of the radial velocity component of the flow. The vertical profiles of $\overline{U_\theta}$ corresponding to different lid angular velocities are shown in figure \ref{fig:Vtheta}. They were obtained by horizontally averaging the horizontal component of the velocity vectors from PIV measurements in a vertical strip at $R$. Figure \ref{fig:Vr} shows $\overline{U_r}$ at different radial distances, including $R$, at a height of 0.5 cm and 7 cm above the basal cooling plate. The radial component of the velocity vectors from PIV measurements in a horizontal sector at these heights were averaged to obtain these profiles of $\overline{U_r}$. 

The $\overline{U_\theta}$ plots show the presence of a thin boundary layer near the bottom plate. Above this boundary layer, there is a core region with uniform $\overline{U_\theta}$ that extends almost to the top of the fluid column. This uniform core can therefore be considered to be in solid body rotation. The velocity in the thin boundary layer near the rough underside surface of the lid has been omitted from the profile as it was difficult to obtain accurate measurements of velocity in that thin layer by PIV due to light reflections from the rough grid degrading the quality of the images. The far-field $\overline{U_\theta}$ is 34\% of the lid velocity at the lowest lid $\Rey$ and 53\% of the lid velocity at the highest lid $\Rey$. $\overline{U_r}$ is 3-4 times larger inside the bottom boundary layer than in the interior of the fluid column. The turbulent flow between a rotating disk and a stationary disk has been studied experimentally by \citet{itoh1992} and \citet{cheah1994} and numerically using LES by \citet{andersson2006}. \citet{itoh1992} also report the presence of an inner core in which $\overline{U_\theta}$ is homogeneous. Denoting $K=\frac{\overline{U_\theta}}{\Omega r}$, they found $K$ in the range 31\% to 42\% for local $\Rey$ ($=\Omega r^2/\nu$) from $1.6 \times 10^5$ to $8.8 \times 10^5$, which corresponds to $1.3 \times 10^4$ to $7.1 \times 10^4$ with the definition of $\Rey$ in (\ref{reynolds}). $\overline{U_r}$ in their experiment was directed inwards in the boundary layer near the stationary plate and was zero in the inner core. In our experiments, the larger values of $K$ at $\Rey$ one order of magnitude smaller and the non-zero $\overline{U_r}$ in the inner core can be attributed to the roughness of the top boundary, which affects the circulation in the tank by causing enhanced mixing. 

\begin{figure}
	\centering
	\begin{subfigure}
		\centering
		\includegraphics{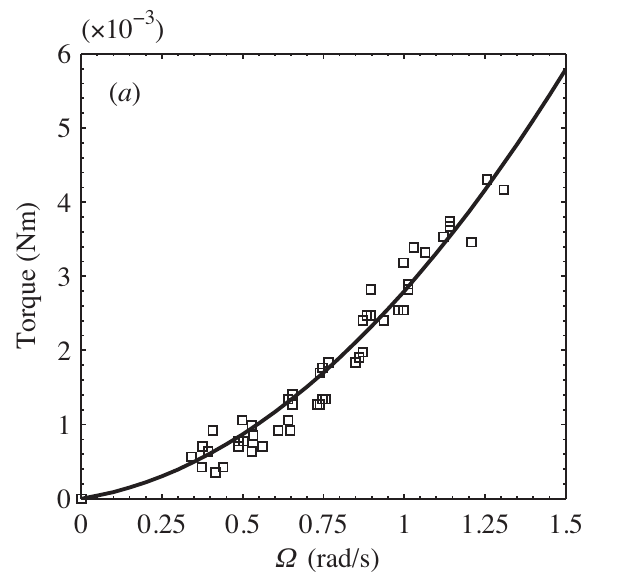}
	\end{subfigure}
	\quad
	\begin{subfigure}
		\centering
		\includegraphics{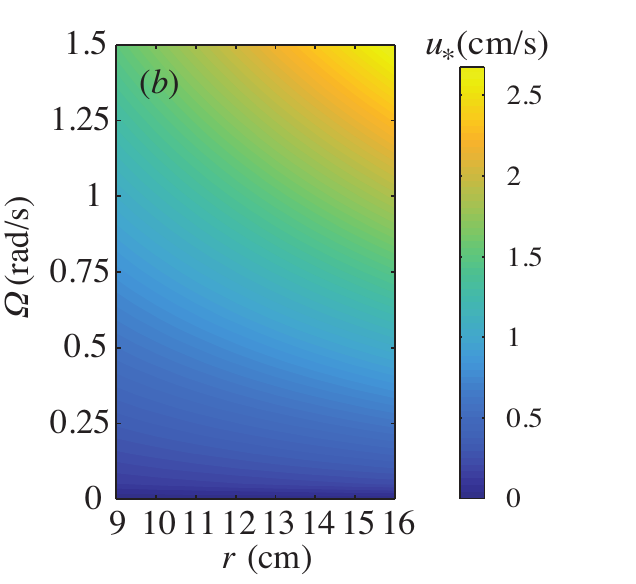}
	\end{subfigure}
	\caption{(\textit{a})Torque on the lid for a water depth of 10 cm: $\square$, measurements; \solidrule: line of best fit. (\textit{b}) friction velocity $u_*$ calculated from torque measurements.  }
	\label{fig:ustar}
\end{figure}

\subsection{Friction Velocity}
The heat balance in (\ref{hbal_p1}) also requires knowledge of the friction velocity $u_\ast$ of the shear-driven flow. Here $u_\ast$ is defined as

\begin{equation}
u_\ast = \sqrt{\frac{\tau(\Omega,r)}{\rho}}
\end{equation}
where $\tau$ is the shear-stress on the lid, which is given by 
\begin{equation}
\tau = C_D \rho_\ell \Omega^2 r^2 
\end{equation}
with $C_D$ being the drag coefficient associated with the lid. Taking $\D F$ to be the incremental change in force along an incremental change in radial distance $\D r$ from the center and $\mathcal{T}$ to be the torque on the lid, $\mathcal{T}$ and $\D F$ are related to $C_D$ by
\begin{equation}
\D F = C_D \rho \Omega^2 r^2 (2\pi r \D r) \\
\end{equation}
and
\begin{equation}
\mathcal{T} = \int_0^R r \ \D F, \\
\end{equation}
so that
\begin{equation}
\mathcal{T} = \frac{2}{5} \pi C_D \rho \Omega^2 R^5.
\end{equation}
Hence $u_\ast$ can be determined from $\mathcal{T}$ according to the relation
\begin{equation}
u_\ast = \sqrt{\frac{5 \mathcal{T} }{2 \pi \rho_\ell R^5}} r
\label{ustar-torque}
\end{equation}

The torque on the lid for different angular velocities of rotation is shown in figure \ref{fig:ustar}. The line of best fit is a weighted-by-value two-parameter polynomial. The friction velocity derived from the torque measurements using (\ref{ustar-torque}) is shown in figure \ref{fig:ustar}.

\section{Model Comparison}\label{sec:modelComp}

\subsection{Ice thickness}
Our heat balance for the ice--water interface can be integrated in time to model the evolution of ice thickness in the experiments conducted with (\ref{hbal_p1}) used for Phase 1 and (\ref{hbal_p2}) used for Phase 2. In what follows, the modelled ice thickness is denoted by $h_m$. The values listed in table \ref{tab:properties} for the properties of liquid water and solid ice are used in the integration. $\Delta T_s$ and $\Delta T_\ell$ in the heat balance are calculated in the following way:  

\begin{figure}
	\centerline{\includegraphics{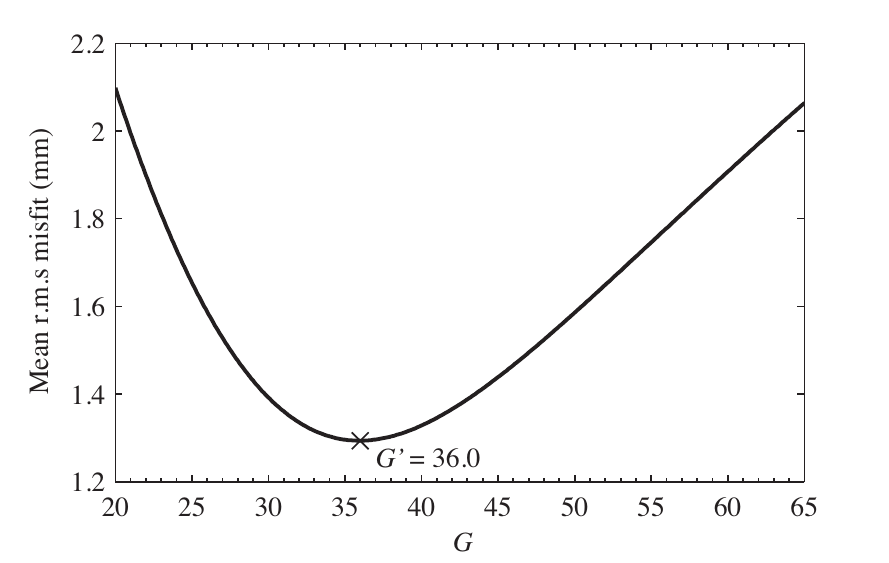}}
	\caption{Misfit to (\ref{hbal_p2}) versus $G$}
	\label{fig:NewG}
\end{figure}

\begin{equation}
\Delta T_s = \left\{
\begin{array}{ll}
T_f - T_A, & \textrm{when ice is present} \\ [2pt]
0, & \textrm{when there is no ice}
\end{array} \right. 
\end{equation}

\begin{equation}
\Delta T_\ell = \left\{
\begin{array}{ll}
T_G - T_f, & \textrm{when ice is present} \\ [2pt]
T_G - T_A, & \textrm{when there is no ice,}
\end{array} \right. 
\end{equation}
where $T_A$ refers to temperature measurements at thermistor A, which is located in a small hole in the basal cooling plate, and $T_G$ refers to temperature measurements at thermistor G, which is located 5.25 cm above the basal cooling plate. The fluid velocity in the far-field, $U_\infty$, is determined using the measurements of $\overline{U_\theta}$ and $\overline{U_r}$ interpolated at the angular velocity of the lid at a height $z=7$ cm:

\begin{equation}
U_\infty = \sqrt{ \overline{U_\theta (z)}^2 + \overline{U_r (z)}^2 }, \ z=7 \ \textrm{cm}.
\end{equation}
The friction velocity $u_\ast$, which is used in calculating the coefficient of friction $c_f$ defined in (\ref{cf}), is determined from the calibration shown in figure \ref{fig:ustar}. The equations are integrated by a second-order Runge-Kutta method, with the initial condition for $h_m$ being zero. Because the temperature measurements were taken at intervals of 5 s, the time-step for integration is also 5 s. 

\begin{figure}
	\centering
	\begin{subfigure}
		\centering
		\includegraphics{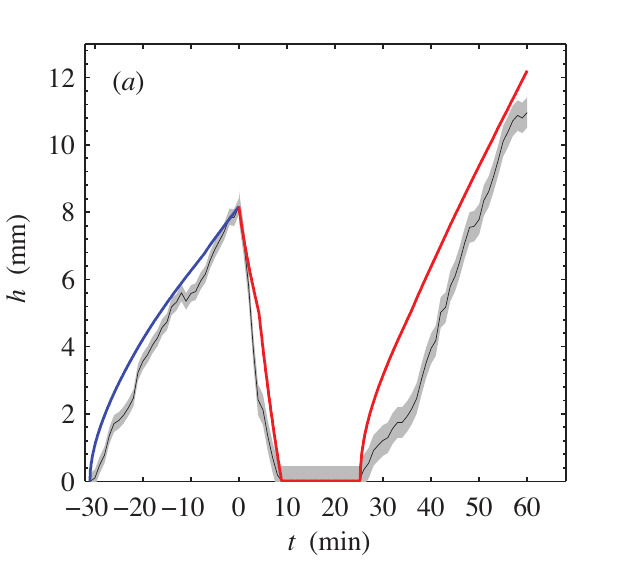}
	\end{subfigure}
	\quad
	\begin{subfigure}
		\centering
		\includegraphics{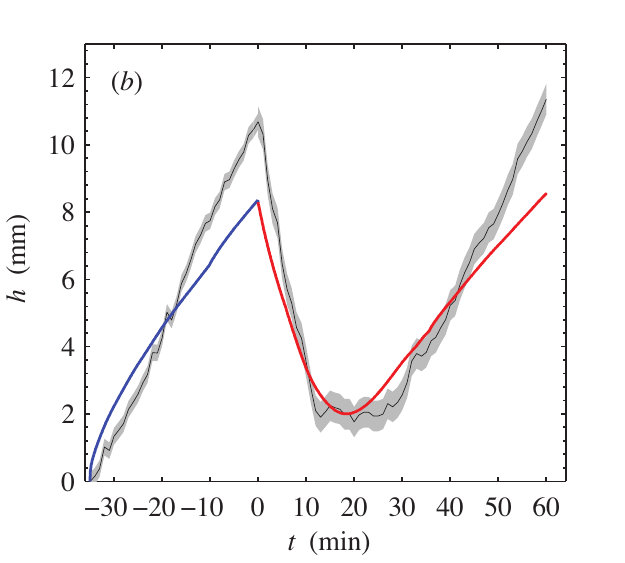}
	\end{subfigure}
	
	\begin{subfigure}
		\centering
		\includegraphics{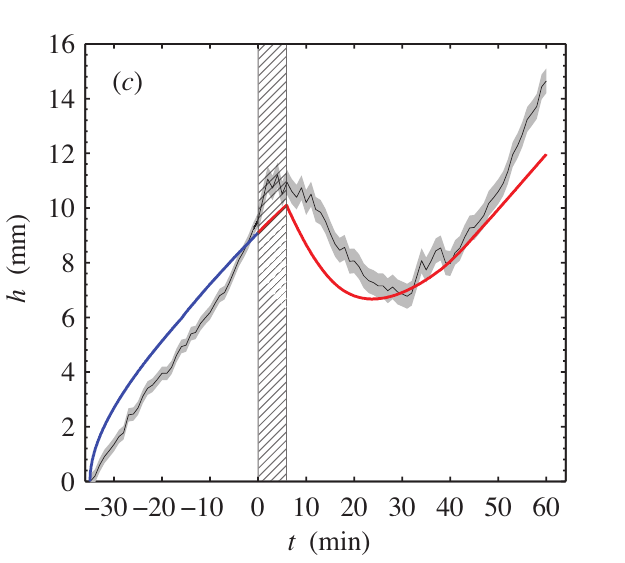}
	\end{subfigure}	
	\quad
	\begin{subfigure}
		\centering
		\includegraphics{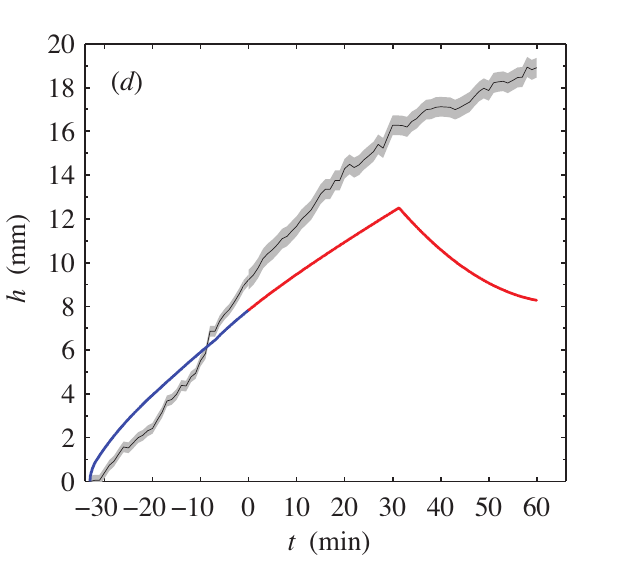}
	\end{subfigure}	
	
	\caption{Comparison of $h_e$ (thinner solid line with shaded error region) with $h_e$ (thicker solid line) from (\textit{a}) Experiment 8, (\textit{b}) Experiment 5, (\textit{c}) Experiment 3, and (\textit{d}) Experiment 1. For $t<0$, the error in $h_e$ is 0.5 mm and for $t>0$, it is 0.9 mm. The hatched region in (\textit{c}) is discussed in the text. (Colour online) For the $h_m$ plots, the blue portion corresponds to ice evolution in Phase 1  while the red portion corresponds ice evolution in Phase 2. }
	\label{fig:ExpModelComparison}
\end{figure}

In Phase 2, the heat flux $q_T$ from the turbulent flow at the ice--water interface depends on the coefficient $G$. For the $\Pran$ of water at 0 \textdegree C, which is listed in table \ref{tab:dimensionless}, $G$ becomes 62.7. We denote this value by $G_0$. The expression for $G$ is an empirical expression derived for a turbulent boundary layer in air over a perfectly flat plate \citep{white1974}. Using a range of values of $G$, including $G_0$, we evaluate $h_m$ during Phase 2 of Experiments 2-10. We also calculate the root mean square (r.m.s) difference $\Delta h_{\rm{RMS}}$ between $h_e$ and $h_m$ at the corresponding times. The omission of Experiment 1 from this comparison will be explained when interpreting figure \ref{fig:ExpModelComparison}(\textit{d}). The mean of $\Delta h_{\rm{RMS}}$ for the range of values of $G$ considered is plotted in figure \ref{fig:NewG}. Its minimum occurs when $G=36.0$. The heat flux from the turbulent layer at the ice--water interface is therefore more closely approximated using this value of $G$, which will be denoted by $G'$. The fact that $G'$ is smaller than $G_0$ indicates that heat transfer from a turbulent flow at an ice--water interface is more efficient than at a flat plate. This enhanced heat transfer can be attributed to the ice surface not being uniformly smooth, especially during melting when it develops a wavy profile, since a rough surface has a greater surface area than a flat surface and hence allows for greater heat transfer. With the new value $G'$, the heat transfer coefficient $C_h$ from the turbulent flow at the ice--water interface given in (\ref{qt}) is related to ($u_\ast/U_\infty$) by the power law
\begin{equation}
C_h = 0.028 \bigg(\frac{u_\ast}{U_\infty} \bigg )^{1.09}.
\end{equation}

\begin{figure}
	\centerline{\includegraphics{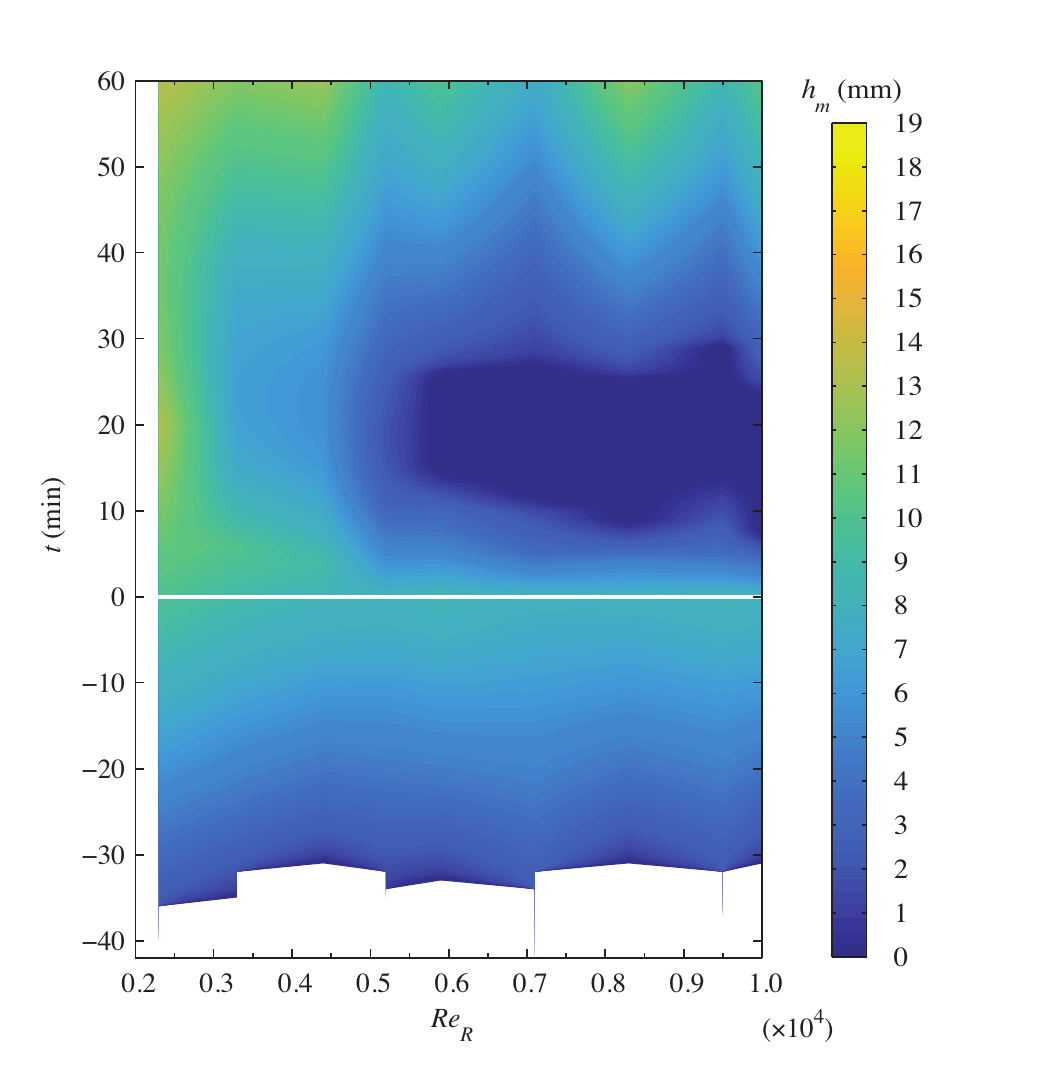}}
	\caption{Contour plot of $h_m$ for $\Rey_R$ corresponding to Experiments 2--10. The white line represents the onset of turbulent mixing. }
	\label{fig:ModelContourThickness}
\end{figure}

 $G'$ is substituted in (\ref{hbal_p2}) to calculate $h_m$ for Phase 2 of Experiments 1--10. Figure \ref{fig:ExpModelComparison} shows the comparison between $h_e$ and $h_m$ in Experiments 8, 5, 3, and 1. There is good agreement between $h_e$ and $h_m$ in Experiments 8, 5, and 3 but not in Phase 2 of Experiment 1. After the onset of turbulent mixing in Experiment 1, a stratified layer remained between the turbulent layer and the ice--water interface up to $t=33$ min. This was evidenced by the behavior of dye inserted into the turbulent layer, with the observation that a clear, stratified layer over the ice--water interface prevented the dyed turbulent layer from reaching the ice surface. Accordingly, the ice growth in that time interval is modelled using (\ref{hbal_p1}). For this case, the measured rate of ice growth is larger than predicted, a difference which occurs because the stratified layer over the ice--water interface inhibits heat transfer from the liquid far-field. Ice grows below the turbulent layer from $t=0$ to $t=33$ min, at which time it reaches the turbulence. For $t>33$ min, $h_m$ is modelled using (\ref{hbal_p2}). The model predicts melting whereas the experimental measurements indicate attenuated growth. The disagreement between $h_e$ and $h_m$ in Experiment 1 shows the limitation of our model when a stratified layer over the ice--water interface persists below the turbulent layer for a long time. This phenomenon occurs at the low end of the range of $\Rey_R$ we investigate, where the applied shear stress from the lid is low, and consequently turbulence is too weak to erode the stratified layer quickly. A modified heat transfer law based on new experiments is needed to model ice thickness in this case. 

\begin{figure}
	\centerline{\includegraphics{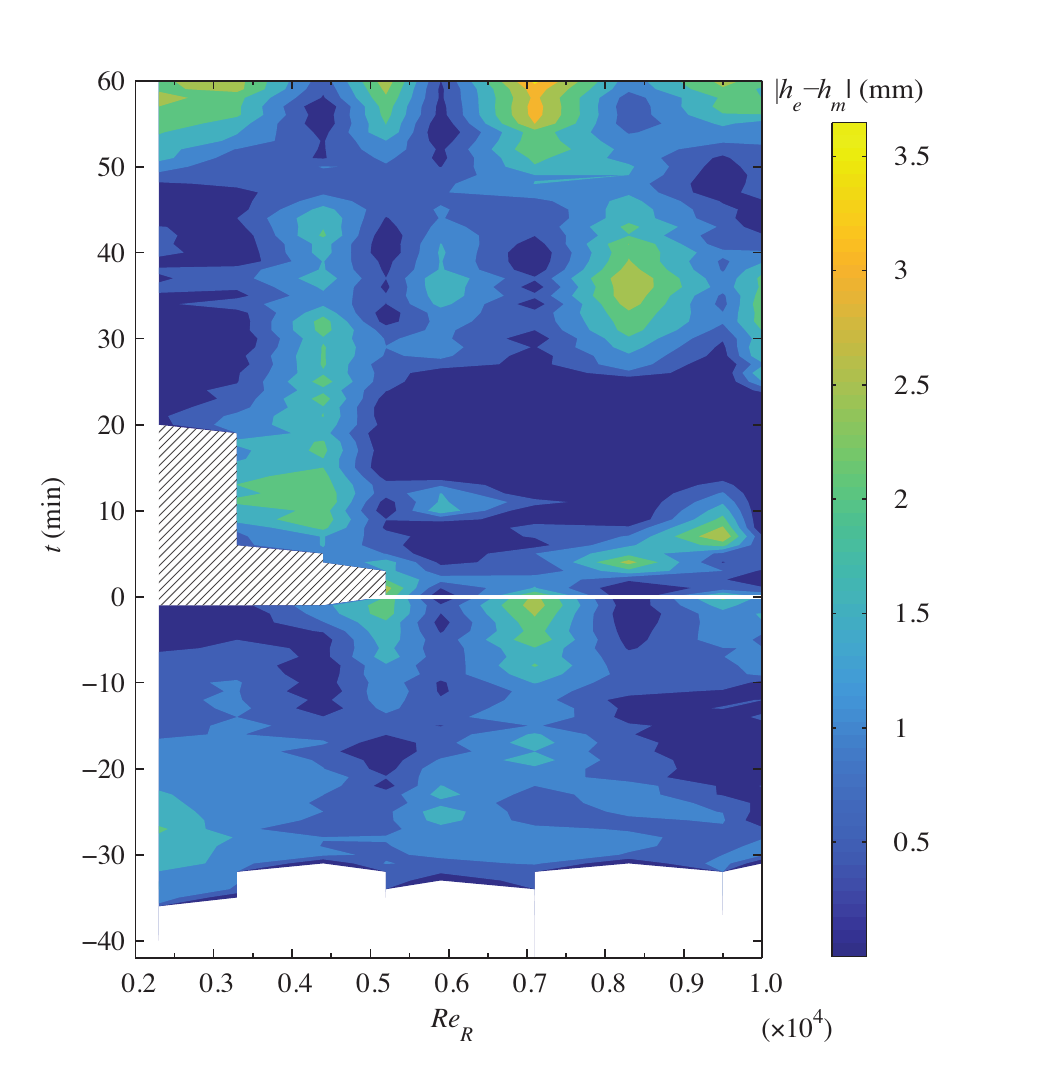}}
	\caption{Contour plot of absolute difference between $h_e$ and $h_m$ for Experiments 2 - 10.}
	\label{fig:Misfit}
\end{figure}

A stratified layer is present in Experiments 2--4 for a short time interval following the onset of turbulent mixing. Although the theoretical model given by (\ref{hbal_p1}) is incomplete for this configuration, we use it to approximate $h_m$ until the time when the turbulent layer comes into contact with the ice. $h_m$ is consistently lower than $h_e$ in that time interval, as shown in the hatched region of figure \ref{fig:ExpModelComparison}(c) for Experiment 3. In the determination of $G'$ previously discussed, $h_e$ and $h_m$ from time intervals when a stratified layer was present were not used. 

The contour plot of $h_m$ calculated using the theoretical model with $G'$ for Experiments 2--10 is shown in figure \ref{fig:ModelContourThickness}. Experiment 1 is omitted because it is a case for which our model is not valid. The absolute difference between $h_e$ from figure \ref{fig:ContourThickness} and $h_m$ at corresponding times is shown in figure \ref{fig:Misfit}. The absolute difference is generally close to the error margin of 0.5 mm for Phase 1 and 0.9 mm for Phase 2, which indicates good agreement between the model and the experiment. The hatched area in the left side of figure \ref{fig:Misfit} correspond to time intervals when a stratified layer was present during Phase 2 of the experiments. The absolute difference between $h_e$ and $h_m$ in these intervals was omitted from the contour plot because the model does not apply correctly. 

We did not observe evidence of the radial component of the flow near the bottom of the tank during turbulent mixing affecting the evolution of ice in our set of experiments. The radial component of the flow is stronger at higher $\Rey_R$. In all experiments in which there is transient melting, the rate at which the ice melts always increases with distance from the center of the tank. During the subsequent re-freezing, the rate of ice growth is always uniform at all radial distances. These observations suggest that the far-field flow has a stronger influence on the evolution of ice than the radial flow near the bottom of the tank. 

\subsection{Liquid temperature in Phase 2}
During Phase 2 of the experiments, turbulent heat transfer is the main mechanism of heat transfer in the liquid. As a result of turbulent mixing, the temperature in the liquid is homogeneous. The enthalpy balance in a control volume in the liquid far-field is 
\begin{equation}
\frac{\D E}{\D t} = q_T.
\end{equation} 
This yields the following expression for the evolution of the homogeneous temperature $T_\ell$ of the liquid:
\begin{equation}
T_\ell(t) = (T_{\ell,0} - T_f) \textrm{e}^{\big ( \frac{-U_\infty C_h}{D-h(t)} \big ) t} + T_f 
\label{tl_model}
\end{equation}
where $T_{\ell,0}$ is the temperature of the liquid at $t=0$ (onset of turbulent mixing).

\begin{figure}
	\centerline{\includegraphics{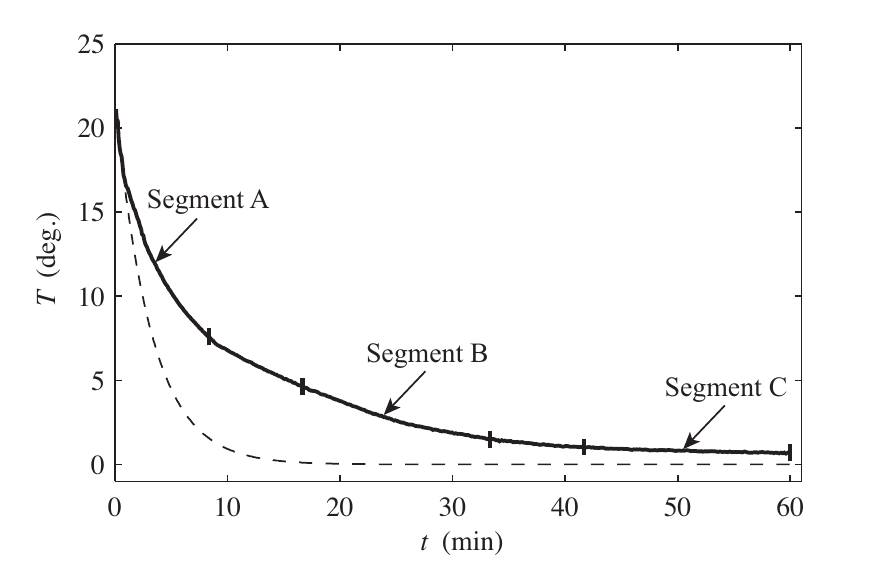}}
	\caption{Turbulent liquid layer temperature evolution from Experiment 9: \solidrule, measured temperature; \protect\dashedrule , modelled temperature.}
	\label{fig:LiquidLayerTemp}
\end{figure}

 Figure \ref{fig:LiquidLayerTemp} shows the measured $T_\ell$ in Phase 2 of a typical experiment. Since the heat transfer coefficient $C_h$ depends on the coefficient $G$, another approach to finding $G$ is by fitting the experimental measurements of the temperature evolution of the turbulently-mixed liquid to (\ref{tl_model}). The values of $G$ obtained from this method are 108, 183, and 620 in the segments A, B, and C respectively shown in table \ref{fig:LiquidLayerTemp} for the typical experiment considered. This method yields values of $G$ that are generally unstable and is thus not pursued. Determining $G$ using (\ref{hbal_p2}) is preferred because measurements of changes in the different components that make up the physical system in the experiment are required, imposing more constraints on $G$. The modelled temperature using the value $G'=36.0$ previously obtained from that approach is also shown in figure \ref{fig:LiquidLayerTemp}. There is poor agreement between the modelled temperature and the measured temperature. The liquid temperature evolution is not truly exponential, however, which means that it does not actually follow the theoretically-predicted trend of (\ref{tl_model}).

\section{Discussion}\label{sec:discus}
\subsection{Application to observations under Pine Island Glacier Ice Shelf}
Pine Island Glacier Ice Shelf is a 40-km-long, 20-km-wide ice shelf in Amundsen Sea off West Antarctica. An investigation involving the deployment of autonomous underwater vehicles in its underlying ocean cavity showed that the basal surface of the ice shelf is experiencing rapid melting, probably due to shoaling Circumpolar Deep Water and intrusion of warmer water under the ice \citep{jenkins2010observations}.

\citet{stanton2013} reported in situ measurements of the basal melt rate and ocean boundary layer properties from a site in the center of Pine Island Glacier Ice Shelf where a hole was drilled vertically from the surface to access the water underneath. We use the measurements, which are listed in table \ref{tab:observations}, to test our heat transfer model from (\ref{hbal_p2}) with $G=G'$. The boundary layer depth and density at the site were obtained from CTD profiling. The departure from freezing, mean current velocity, mean friction velocity, and local melting rate were measured using a flux package installed at an initial distance of 2.3 m below the ice shelf base. The range of values listed for the departure from freezing and mean current velocity are for a 35-day period. The mean friction velocity was constant in that period. The local melt rate is from a fit through measurements from days 5-35 and is equivalent to 14 m per year.

\begin{table}
	\begin{center}
		\def~{\hphantom{0}}
		\begin{tabular}{lccc}
			Property  & In (\ref{hbal_p2}) & Units & Value \\
			Ice shelf thickness & $h$ & m & 460  \\
			Boundary layer density & $\rho_\ell$ & kgm$^{-3}$ & 27.22 - 27.42  \\
			Departure from freezing & $\Delta T_\ell$ & K & 1.35 - 1.42 \\
			Mean current velocity & $U_\infty$ & ms$^{-1}$ & 0.11 - 0.15 \\
			Mean friction velocity & $u_\ast$ & ms$^{-1}$ & 0.0086 \\
			Local melting rate & $-\D h/\D t$ & m per day & 0.039 \\
			\\
			Buoyancy frequency, $N = \sqrt{(-g/\rho)(\partial \rho/{\tiny }\partial z)}$ & & s$^{-1}$ & 0.021 \\
			Vertical velocity gradient, $\frac{\partial u}{\partial z}$ & & s$^{-1}$ & 0.057
		\end{tabular}
		\caption{Measurements of the ocean boundary layer properties from \citet{stanton2013} and our estimates of the buoyancy frequency $N$ and vertical velocity gradient $\D u/\D z$ calculated from the measured properties.}
		\label{tab:observations}
	\end{center}
\end{table}

We substitute $h$, $u_\ast$, and the medians of the range of values of $\rho_\ell$, $\Delta T_\ell$, and $U_\infty$ from table \ref{tab:observations}, $\rho_s$, $k_s$, and $L$ from table \ref{tab:properties}, a typical value of  $c_\ell = 4.00 \times 10^3$ Jkg$^{-1}$K for sea water, and a typical value of $\Delta T_s = 25$ K for an ice shelf in (\ref{hbal_p2}). This yields a corresponding predicted melt rate $-\D h/\D t$ of 98 m per year. The fact that our model over-predicts the observed melt rate can be explained by the observed $\Delta T_\ell$ being across a thick stratified boundary layer, a case for which our model is not valid. 

Consideration of the effect of stable stratification suggests a possible explanation for this discrepancy. One way of estimating the decrease
is through the gradient Richardson number $Ri_g$, which is a measure of the relative strength of the stabilizing effect from density stratification compared to the destabilizing effect caused by turbulent shear. It is expressed as 
\begin{equation}
Ri_g = \frac{N^2}{(\partial u/\partial z)^2}
\end{equation}
where $N$ is the buoyancy frequency. For the measured flow under Pine Island Glacier Ice Shelf, $Ri_g = 0.14$ using the values of $N$ and $\partial u/\partial z$ estimated from the boundary layer properties listed in table \ref{tab:observations}. According to \citet{galperin2007critical}, the associated eddy diffusivity is about 1/2 of the eddy diffusivity of a flow in which there is no stratification and $Ri_g$ is zero. Stratification thus inhibits heat exchange between the ocean and ice shelf base. Alternatively, following Monin-Obukhov similarity theory, the ratio of the measured shear to the shear from a theoretical logarithmic velocity profile at a distance $z$ from the basal surface is
\begin{equation}
\bigg ( \frac{\partial u}{\partial z} \bigg ) \bigg ( \frac{\kappa z}{u_\ast} \bigg ) = \phi,
\end{equation}
where $\kappa$ is the von K\'arm\'an constant, taken to be 0.4. At the measurement location, $z=2.3$ m and $\phi \simeq 6$. The actual momentum flux within the boundary layer is thus 1/6 of its mean gradient. Assuming the same behavior for heat flux, the turbulent transfer of heat would also reduced by a factor of 6. Considering these two approaches of estimating the effect of stratification, the actual melt rate should lie between 1/6 and 1/2 of the predicted value from our model. With this correction applied, the resulting values lie between 16 and 49 m per year and overlap the observed range of 14--24 m per year from \citet{stanton2013}. Unfortunately the uncertainties in stratification-dependence of our model and in observational melt rate make it impossible to say whether our result is more accurate with $G'$ rather than $G_0$ used in (\ref{hbal_p2}). (Using $G_0$ gives a predicted melt rate which is about 2/3 the value from using $G'$.)

 Our predicted value of the melt rate at the measurement site is an upper limit and corresponds to the case where turbulent warm water flow comes in direct contact with the ice interface. The measurements by \citet{stanton2013} were taken in the crest of a channel at the base of Pine Island Glacier Ice Shelf. The ice shelf base is laterally heterogeneous. A thick buoyant layer is often trapped in the crest of channels at the basal surfaces of ice shelves and in areas outside the channels, the buoyant layer is much thinner \citep{gladish2012}. Our heat transfer model is more applicable to these areas. \citet{jenkins2010observations} reported, from autosub observations, that the basal melting rate in the central part of Pine Island Glacier Ice Shelf is greater than 50 m per year. The mechanism responsible for basal melting of the ice shelf over a larger area may thus be closer to the mechanism we consider in our simple one-dimensional model.

\subsection{Implication for interpretation of field measurements}
 The poor agreement between the modelled temperature evolution of the liquid and the measured temperature in our experiments suggests that the oceanic boundary layer near the ice-ocean interface cannot be assumed to be a calorimeter if it contains a large amount of meltwater. According to our experiments, using only the temperature measurements of the ocean water underneath the ice shelf or at the ice shelf front is ill-suited for estimating ice shelf thickness evolution. In field studies, measurements in the solid ice and the liquid water at the ice-ocean interface are needed to develop a constrained model for predicting basal melting of ice shelves in response to changes in ocean properties.

\begin{figure}
	\centerline{\includegraphics{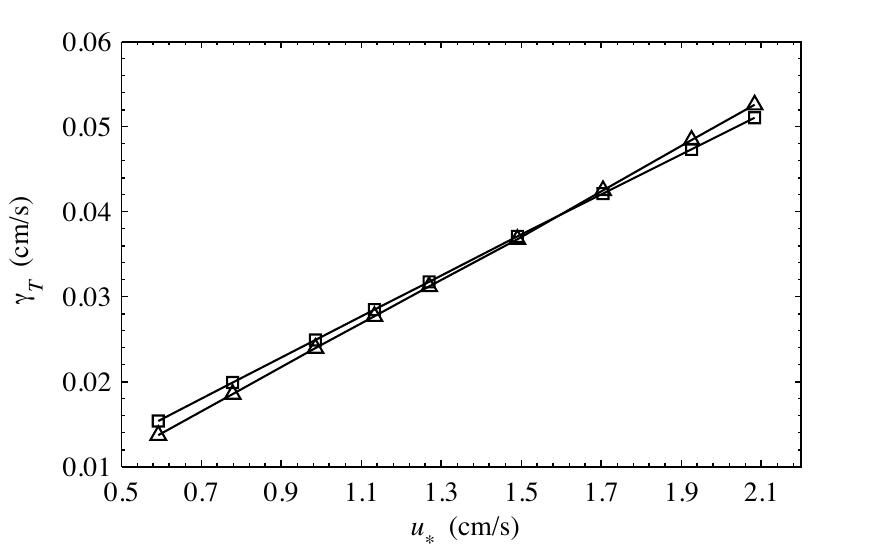}}
	\caption{Thermal exchange velocity $\gamma_T$ corresponding to the $u_\ast$ from our experiments: $\triangle$  from (\ref{gamma_T1}) and $\square$ from (\ref{gamma_T2}) with $M=-2.4$.}
	\label{fig:gamma_comparison}
\end{figure}

\subsection{Comparison of $\gamma_T$ from our model with $\gamma_T$ from \citet{jenkins1991}}
The parameterization of the thermal exchange velocity across the boundary layer in our model is 
\begin{equation}
\gamma_T = U_\infty C_h = U_\infty \bigg ( \frac{c_f/2}{1 + G'\sqrt{c_f/2}} \bigg ).
\label{gamma_T1}
\end{equation}
\citet{jenkins1991} expressed $\gamma_T$ as
\begin{equation}
\gamma_T = \frac{C_d^{1/2} U_\infty}{2.12 \ln(C_d^{1/2} \Rey) + 12.5 \Pran ^{2/3} - 8.68}
\end{equation}
where $C_d$ is a dimensionless drag coefficient given by 
\begin{equation}
C_d = \frac{u_\ast^2}{U_\infty^2}.
\end{equation}
The two expressions for $\gamma_T$ are essentially equivalent, being related by the Prandtl-Nikuradse skin friction law \citep{kader1972}. The term $(12.5 \Pran^{2/3} - 8.68)$ in the expression of \citet{jenkins1991} is a constant which we denote by $M$:
\begin{equation}
\gamma_T = \frac{C_d^{1/2} U_\infty}{2.12 \ln(C_d^{1/2} \Rey) + M}.
\label{gamma_T2}
\end{equation}
M evaluates to 62.5 using the $\Pran$ listed in table \ref{tab:properties}. We calculate $\gamma_T$ using (\ref{gamma_T1}) and (\ref{gamma_T2}) for the measured values of $u_\ast$ and $U_\infty$ measured from our experiments. The r.m.s. difference between the two sets of $\gamma_T$ values is minimum if $M$ is adjusted to --2.4. The comparison is shown in figure \ref{fig:gamma_comparison}.

\section{Summary}\label{sec:summary}
We have conducted experiments on the melting of ice in a turbulent shear flow that transports warm water to the ice--water interface. A modified heat transfer law, originally derived for turbulent flow over a flat plate and which depends on the friction velocity of the flow, allows us to model the evolution of the ice thickness correctly. Our experiments have dynamic similarity with the geophysical system of the ocean cavity beneath an ice shelf through the Rossby number and thermodynamic similarity through the Stefan number. Although our experiments do not include the effect of geometry of the ocean cavity and the effect of salinity on the freezing point of ice in the ocean, they reveal the mechanisms through which warm water transport to an ice shelf's basal surface accelerates basal melting. Through this study, we propose an experimentally-constrained expression for the thermal exchange velocity $\gamma_T$ in ice-ocean interaction. 
\\
\par
We thank Yuanchao Li and Huang Chen for help in conducting the PIV measurements. The experiments were financially supported by NSF grant EAR-110371.

\bibliographystyle{jfm}
\bibliography{myReferences}

\begin{thebibliography}{41}
\expandafter\ifx\csname natexlab\endcsname\relax\def\natexlab#1{#1}\fi
\def\au#1{#1} \def\ed#1{#1} \def\yr#1{#1}\def\at#1{#1}\def\jt#1{\textit{#1}}
  \def\bt#1{#1}\def\bvol#1{\textbf{#1}} \def\vol#1{#1} \def\pg#1{#1}
  \def\publ#1{#1}\def\arxiv#1{#1}\def\org#1{#1}\def\st#1{\textit{#1}}

\bibitem[Andersson \& Lygren(2006)]{andersson2006}
{\sc \au{Andersson, H.~I} \& \au{Lygren, M.}} \yr{2006}  \at{{LES} of open
  rotor--stator flow}.  \jt{Int. J. Heat Fluid Flow}  \bvol{27}~(4),
  \pg{551--557}.

\bibitem[Boger \& Westwater(1967)]{boger1967effect}
{\sc \au{Boger, D.~V.} \& \au{Westwater, J.~W.}} \yr{1967}  \at{Effect of
  buoyancy on the melting and freezing process}.  \jt{J. Heat Transfer}
  \bvol{89}~(1),  \pg{81--89}.

\bibitem[Brisbourne {\em et~al.\/}(2014)Brisbourne, Smith, King, Nicholls,
  Holland \& Makinson]{brisbourne2014}
{\sc \au{Brisbourne, A.~M.}, \au{Smith, A.~M.}, \au{King, E.~C.}, \au{Nicholls,
  K.~W.}, \au{Holland, P.~R.} \& \au{Makinson, K.}} \yr{2014}  \at{Seabed
  topography beneath {Larsen C Ice Shelf} from seismic soundings}.
  \jt{Cryosphere}  \bvol{8},  \pg{1--13}.

\bibitem[Broecker {\em et~al.\/}(1998)Broecker, Peacock, Walker, Weiss,
  Fahrbach, Schr{\"o}der, Mikolajewicz, Heinze, Key, Peng \&
  Rubin]{broecker1998}
{\sc \au{Broecker, W.~S.}, \au{Peacock, S.~L.}, \au{Walker, S.}, \au{Weiss,
  R.}, \au{Fahrbach, E.}, \au{Schr{\"o}der, M.}, \au{Mikolajewicz, U.},
  \au{Heinze, C.}, \au{Key, R.}, \au{Peng, T.-H.} \& \au{Rubin, S.}} \yr{1998}
  \at{How much deep water is formed in the {Southern Ocean}?}  \jt{J. Geophys.
  Res.}  \bvol{103}~(C8),  \pg{15833--15843}.

\bibitem[Cheah {\em et~al.\/}(1994)Cheah, Iacovides, Jackson, Ji \&
  Launder]{cheah1994}
{\sc \au{Cheah, S.~C.}, \au{Iacovides, H.}, \au{Jackson, D.~C.}, \au{Ji, H.} \&
  \au{Launder, B.~E.}} \yr{1994}  \at{Experimental investigation of enclosed
  rotor-stator disk flows}.  \jt{Exp. Therm. Fluid Sci.}  \bvol{9}~(4),
  \pg{445--455}.

\bibitem[Dallaston {\em et~al.\/}(2015)Dallaston, Hewitt \&
  Wells]{dallaston2015channelization}
{\sc \au{Dallaston, M.~C.}, \au{Hewitt, I.~J.} \& \au{Wells, A.~J.}} \yr{2015}
  \at{Channelization of plumes beneath ice shelves}.  \jt{J. Fluid Mech.}
  \bvol{785},  \pg{109--134}.

\bibitem[Dansereau {\em et~al.\/}(2014)Dansereau, Heimbach \&
  Losch]{dansereau2014}
{\sc \au{Dansereau, V.}, \au{Heimbach, P.} \& \au{Losch, M.}} \yr{2014}
  \at{Simulation of subice shelf melt rates in a general circulation model:
  Velocity-dependent transfer and the role of friction}.  \jt{J. Geophys. Res.
  Oceans}  \bvol{119},  \pg{1765--1790}.

\bibitem[Feldmann \& Levermann(2015)]{feldmann2015collapse}
{\sc \au{Feldmann, J.} \& \au{Levermann, A.}} \yr{2015}  \at{Collapse of the
  {W}est {A}ntarctic {I}ce {S}heet after local destabilization of the
  {A}mundsen {B}asin}.  \jt{Proc. Natl. Acad. Sci. USA}  \pg{p. 201512482}.

\bibitem[Galperin {\em et~al.\/}(2007)Galperin, Sukoriansky \&
  Anderson]{galperin2007critical}
{\sc \au{Galperin, B.}, \au{Sukoriansky, S.} \& \au{Anderson, P.~S.}} \yr{2007}
   \at{On the critical {R}ichardson number in stably stratified turbulence}.
  \jt{Atmos. Sci. Lett.}  \bvol{8}~(3),  \pg{65--69}.

\bibitem[Gilpin {\em et~al.\/}(1980)Gilpin, Hirata \& Cheng]{gilpin1980wave}
{\sc \au{Gilpin, R.R.}, \au{Hirata, T.} \& \au{Cheng, K.~C.}} \yr{1980}
  \at{Wave formation and heat transfer at an ice-water interface in the
  presence of a turbulent flow}.  \jt{J.~Fluid Mech.}  \bvol{99},
  \pg{619--640}.

\bibitem[Gladish {\em et~al.\/}(2012)Gladish, Holland, Holland \&
  Price]{gladish2012}
{\sc \au{Gladish, C.~V.}, \au{Holland, D.~M.}, \au{Holland, P.~R.} \&
  \au{Price, S.~F.}} \yr{2012}  \at{Ice-shelf basal channels in a coupled
  ice/ocean model}.  \jt{J. Glaciol.}  \bvol{58}~(212),  \pg{1227--1244}.

\bibitem[Haynes(2015)]{haynes2015crc}
{\sc \au{Haynes, W.~M.}} \yr{2015} {\em CRC Handbook of Chemistry and Physics:
  A Ready-Reference Book of Chemical and Physical Data\/}.  \publ{CRC}.

\bibitem[Hellmer \& Olbers(1989)]{hellmer1989}
{\sc \au{Hellmer, H.~H.} \& \au{Olbers, D.~J.}} \yr{1989}  \at{A
  two-dimensional model for the thermohaline circulation under an ice shelf}.
  \jt{Antarc. Sci.}  \bvol{1}~(4),  \pg{325--336}.

\bibitem[Holland \& Jenkins(1999)]{holland1999}
{\sc \au{Holland, D.~M.} \& \au{Jenkins, A.}} \yr{1999}  \at{Modeling
  thermodynamic ice-ocean interactions at the base of an ice shelf}.  \jt{J.
  Phys. Oceanogr.}  \bvol{29}~(8),  \pg{1787--1800}.

\bibitem[Holland \& Feltham(2006)]{holland2006}
{\sc \au{Holland, P.~R.} \& \au{Feltham, D.~L.}} \yr{2006}  \at{The effects of
  rotation and ice shelf topography on frazil-laden ice shelf water plumes}.
  \jt{J. Phys. Oceanogr.}  \bvol{36}~(12),  \pg{2312--2327}.

\bibitem[Hooke(2005)]{hooke2005}
{\sc \au{Hooke, R.~L.}} \yr{2005} {\em {Principles of Glacier Mechanics}\/}.
  \publ{Cambridge University Press}.

\bibitem[Huppert \& Worster(1985)]{huppert1985}
{\sc \au{Huppert, H.~E.} \& \au{Worster, M.~G.}} \yr{1985}  \at{Dynamic
  solidification of a binary melt}.  \jt{Nature}  \bvol{314},  \pg{703--707}.

\bibitem[IOC {\em et~al.\/}(2010)IOC, SCOR \& IAPSO]{iociapso}
{\sc \au{IOC}, \au{SCOR} \& \au{IAPSO}} \yr{2010} {\em The international
  thermodynamic equation of seawater --- 2010: Calculation and use of
  thermodynamic properties.\/}.  \publ{{I}ntergovernmental {O}ceanographic
  {C}ommission, {M}anuals and {G}uides No. 56, UNESCO (English)}.

\bibitem[Itoh {\em et~al.\/}(1992)Itoh, Yamada, Imao \& Gonda]{itoh1992}
{\sc \au{Itoh, M.}, \au{Yamada, Y.}, \au{Imao, S.} \& \au{Gonda, M.}} \yr{1992}
   \at{Experiments on turbulent flow due to an enclosed rotating disk}.
  \jt{Exp. Therm. Fluid Sci.}  \bvol{5}~(3),  \pg{359--368}.

\bibitem[Jacobs {\em et~al.\/}(2011)Jacobs, Jenkins, Giulivi \&
  Dutrieux]{jacobs2011}
{\sc \au{Jacobs, S.~S.}, \au{Jenkins, A.}, \au{Giulivi, C.~F.} \& \au{Dutrieux,
  P.}} \yr{2011}  \at{Stronger ocean circulation and increased melting under
  {Pine Island Glacier} ice shelf}.  \jt{Nat. Geosci.}  \bvol{4},
  \pg{519--523}.

\bibitem[Jenkins(1991)]{jenkins1991}
{\sc \au{Jenkins, A.}} \yr{1991}  \at{A one-dimensional model of ice
  shelf-ocean interaction}.  \jt{J. Geophys. Res. Oceans}  \bvol{96}~(C11),
  \pg{20671--20677}.

\bibitem[Jenkins {\em et~al.\/}(2010{\natexlab{{\em a\/}}})Jenkins, Dutrieux,
  Jacobs, McPhail, Perrett, Webb \& White]{jenkins2010observations}
{\sc \au{Jenkins, A.}, \au{Dutrieux, P.}, \au{Jacobs, S.~S.}, \au{McPhail,
  S.~D.}, \au{Perrett, J.~R.}, \au{Webb, A.~T.} \& \au{White, D.}}
  \yr{2010{\natexlab{{\em a\/}}}}  \at{Observations beneath {Pine Island
  Glacier} in {West Antarctica} and implications for its retreat}.  \jt{Nat.
  Geosci.}  \bvol{3},  \pg{468--472}.

\bibitem[Jenkins {\em et~al.\/}(2010{\natexlab{{\em b\/}}})Jenkins, Nicholls \&
  Corr]{jenkins2010}
{\sc \au{Jenkins, A.}, \au{Nicholls, K.~W.} \& \au{Corr, H. F.~J.}}
  \yr{2010{\natexlab{{\em b\/}}}}  \at{Observation and parameterization of
  ablation at the base of {Ronne Ice Shelf, Antarctica}}.  \jt{J. Phys.
  Oceanogr.}  \bvol{40}~(10),  \pg{2298--2312}.

\bibitem[Kader \& Yaglom(1972)]{kader1972}
{\sc \au{Kader, B.~A.} \& \au{Yaglom, A.~M.}} \yr{1972}  \at{Heat and mass
  transfer laws for fully turbulent wall flows}.  \jt{Int. J. Heat Mass
  Transfer}  \bvol{15},  \pg{2329--2351}.

\bibitem[Kerr \& McConnochie(2015)]{kerr2015dissolution}
{\sc \au{Kerr, R.~C.} \& \au{McConnochie, C.~D.}} \yr{2015}  \at{Dissolution of
  a vertical solid surface by turbulent compositional convection}.  \jt{J.
  Fluid Mech.}  \bvol{765},  \pg{211--228}.

\bibitem[Little {\em et~al.\/}(2008)Little, Gnanadesikan \&
  Hallberg]{little2008large}
{\sc \au{Little, C.~M.}, \au{Gnanadesikan, A.} \& \au{Hallberg, R.}} \yr{2008}
  \at{Large-scale oceanographic constraints on the distribution of melting and
  freezing under ice shelves}.  \jt{J. Phys. Oceanogr.}  \bvol{38}~(10),
  \pg{2242--2255}.

\bibitem[Marinov {\em et~al.\/}(2008)Marinov, Gnanadesikan, Sarmiento,
  Toggweiler, Follows \& Mignone]{marinov2008}
{\sc \au{Marinov, I.}, \au{Gnanadesikan, A.}, \au{Sarmiento, J.~L.},
  \au{Toggweiler, J.~R.}, \au{Follows, M.} \& \au{Mignone, B.~K.}} \yr{2008}
  \at{Impact of oceanic circulation on biological carbon storage in the ocean
  and atmospheric p{CO}2}.  \jt{Glob. Biogeochem. Cycles}  \bvol{22}.

\bibitem[McPhee {\em et~al.\/}(1987)McPhee, Maykut \& Morison]{mcphee1987}
{\sc \au{McPhee, M.~G.}, \au{Maykut, G.~A.} \& \au{Morison, J.~H.}} \yr{1987}
  \at{Dynamics and thermodynamics of the ice/upper ocean system in the marginal
  ice zone of the greenland sea}.  \jt{J. Geophys. Res. Oceans}
  \bvol{92}~(C7),  \pg{7017--7031}.

\bibitem[Mueller {\em et~al.\/}(2012)Mueller, Padman, Dinniman, Erofeeva,
  Fricker \& King]{mueller2012}
{\sc \au{Mueller, R.~D.}, \au{Padman, L.}, \au{Dinniman, M.~S.}, \au{Erofeeva,
  S.~Y.}, \au{Fricker, H.~A.} \& \au{King, M.~A.}} \yr{2012}  \at{Impact of
  tide-topography interactions on basal melting of {Larsen C Ice Shelf,
  Antarctica}}.  \jt{J. Geophys. Res. Oceans}  \bvol{117}~(C5).

\bibitem[Neufeld \& Wettlaufer(2008)]{neufeld2008experimental}
{\sc \au{Neufeld, J.~A.} \& \au{Wettlaufer, J.~S.}} \yr{2008}  \at{An
  experimental study of shear-enhanced convection in a mushy layer}.  \jt{J.
  Fluid Mech.}  \bvol{612},  \pg{363--385}.

\bibitem[Pritchard {\em et~al.\/}(2012)Pritchard, Ligtenberg, Fricker, Vaughan,
  Van~den Broeke \& Padman]{pritchard2012}
{\sc \au{Pritchard, H.~D.}, \au{Ligtenberg, S. R.~M.}, \au{Fricker, H.~A.},
  \au{Vaughan, D.~G.}, \au{Van~den Broeke, M.~R.} \& \au{Padman, L.}} \yr{2012}
   \at{Antarctic ice-sheet loss driven by basal melting of ice shelves}.
  \jt{Nature}  \bvol{484},  \pg{502--505}.

\bibitem[Scheduikat \& Olbers(1990)]{scheduikat1990}
{\sc \au{Scheduikat, M.} \& \au{Olbers, D.~J.}} \yr{1990}  \at{A
  one-dimensional mixed layer model beneath the {Ross Ice Shelf} with tidally
  induced vertical mixing}.  \jt{Antarc. Sci.}  \bvol{2}~(1),  \pg{29--42}.

\bibitem[Schmidtko {\em et~al.\/}(2014)Schmidtko, Heywood, Thompson \&
  Aoki]{schmidtko2014}
{\sc \au{Schmidtko, S.}, \au{Heywood, K.~J.}, \au{Thompson, A.~F.} \& \au{Aoki,
  S.}} \yr{2014}  \at{Multidecadal warming of {Antarctic} waters}.
  \jt{Science}  \bvol{346},  \pg{1227--1231}.

\bibitem[Stanton {\em et~al.\/}(2013)Stanton, Shaw, Truffer, Corr, Peters,
  Riverman, Bindschadler, Holland \& Anandakrishnan]{stanton2013}
{\sc \au{Stanton, T.~P.}, \au{Shaw, W.~J.}, \au{Truffer, M.}, \au{Corr, H.
  F.~J.}, \au{Peters, L.~E.}, \au{Riverman, K.~L.}, \au{Bindschadler, R.},
  \au{Holland, D.~M.} \& \au{Anandakrishnan, S.}} \yr{2013}  \at{Channelized
  ice melting in the ocean boundary layer beneath {Pine Island Glacier,
  Antarctica}}.  \jt{Science}  \bvol{341},  \pg{1236--1239}.

\bibitem[Stern {\em et~al.\/}(2014)Stern, Holland, Holland, Jenkins \&
  Sommeria]{stern2014effect}
{\sc \au{Stern, A.~A.}, \au{Holland, D.~M.}, \au{Holland, P.~R.}, \au{Jenkins,
  A.} \& \au{Sommeria, J.}} \yr{2014}  \at{The effect of geometry on ice shelf
  ocean cavity ventilation: a laboratory experiment}.  \jt{Exp. Fluids}
  \bvol{55}~(5),  \pg{1--19}.

\bibitem[Thielicke \& Stamhuis(2014)]{thielicke2014pivlab}
{\sc \au{Thielicke, W.} \& \au{Stamhuis, E.~J.}} \yr{2014}
  \at{{PIV}lab--towards user-friendly, affordable and accurate digital particle
  image velocimetry in matlab}.  \jt{J. Open Res. Software}  \bvol{2}~(1),
  \pg{e30}.

\bibitem[Townsend(1964)]{townsend1964natural}
{\sc \au{Townsend, A.~A.}} \yr{1964}  \at{Natural convection in water over an
  ice surface}.  \jt{Quart. J. Roy. Meteor. Soc.}  \bvol{90}~(385),
  \pg{248--259}.

\bibitem[Wagner \& Pru{\ss}(2002)]{wagner2002}
{\sc \au{Wagner, W.} \& \au{Pru{\ss}, A.}} \yr{2002}  \at{The {IAPWS}
  formulation 1995 for the thermodynamic properties of ordinary water substance
  for general and scientific use}.  \jt{J. Phys. Chem. Ref. Data}
  \bvol{31}~(2),  \pg{387--535}.

\bibitem[Wettlaufer {\em et~al.\/}(1997)Wettlaufer, Worster \&
  Huppert]{wettlaufer1997}
{\sc \au{Wettlaufer, J.~S.}, \au{Worster, M.~G.} \& \au{Huppert, H.~E.}}
  \yr{1997}  \at{Natural convection during solidification of an alloy from
  above with application to the evolution of sea ice}.  \jt{J.~Fluid Mech.}
  \bvol{344},  \pg{291--316}.

\bibitem[White(1974)]{white1974}
{\sc \au{White, F.~M.}} \yr{1974} {\em Viscous Fluid Flow\/}.
  \publ{McGraw-Hill}.

\bibitem[Yaglom \& Kader(1974)]{yaglom1974}
{\sc \au{Yaglom, A.~M.} \& \au{Kader, B.~A.}} \yr{1974}  \at{Heat and mass
  transfer between a rough wall and turbulent fluid flow at high {Reynolds} and
  {Peclet} numbers}.  \jt{J. Fluid Mech.}  \bvol{62},  \pg{601--623}.

\end{thebibliography}

\end{document}